\documentclass[a4paper,fleqn,usenatbib]{mnras}

\usepackage[T1]{fontenc}
\usepackage{ae,aecompl}

\usepackage{graphicx}
\usepackage{amsmath}
\usepackage{amssymb}

\title[Asteroseismic measurement of the helium abundance]{Helium abundance in a sample of cool stars: measurements from asteroseismology}

\author[Verma et al.]{Kuldeep Verma,$^{1}$\thanks{E-mail: kuldeep@phys.au.dk (KV)}
Keyuri Raodeo,$^{2}$
Sarbani Basu,$^{3}$
V\'{i}ctor Silva Aguirre,$^{1}$
\newauthor
Anwesh Mazumdar,$^{2}$
Jakob R{\o}rsted Mosumgaard,$^{1}$
Mikkel N.~Lund,$^{4,1}$
\newauthor
and Pritesh Ranadive$^{2}$
\\
$^{1}$Stellar Astrophysics Centre, Department of Physics and Astronomy, Aarhus University, Ny Munkegade 120, DK-8000 Aarhus C, Denmark\\
$^{2}$Homi Bhabha Centre for Science Education, TIFR, V.~N. Purav Marg, Mankhurd, Mumbai 400088, India\\
$^{3}$Astronomy Department, Yale University, P.~O. Box 208101, New Haven, CT 065208101, USA\\
$^{4}$School of Physics and Astronomy, University of Birmingham, Edgbaston, Birmingham, B15 2TT, UK
}

\date{Accepted XXX. Received YYY; in original form ZZZ}
\pubyear{2018}

\begin{document}
\label{firstpage}
\pagerange{\pageref{firstpage}--\pageref{lastpage}}
\maketitle

\begin{abstract}
The structural stratification of a solar-type main sequence star primarily depends on its mass and chemical composition. The surface heavy 
element abundances of the solar-type stars are reasonably well determined using conventional spectroscopy, however the second most abundant element 
helium is not. This is due to the fact that the envelope temperature of such stars is not high enough to excite helium. Since the helium abundance 
of a star affects its structure and subsequent evolution, the uncertainty in the helium abundance of a star makes estimates of its global properties 
(mass, radius, age etc.) uncertain as well. The detections of the signatures of the acoustic glitches from the precisely measured stellar oscillation
frequencies provide an indirect way to estimate the envelope helium content. We use the glitch signature caused by the ionization of helium to 
determine the envelope helium abundance of 38 stars in the {\it Kepler} seismic LEGACY sample. Our results confirm that atomic diffusion does indeed 
take place in solar-type stars. We use the measured surface abundances in combination with the settling predicted by the stellar models to determine 
the initial abundances. The initial helium and metal mass fractions have subsequently been used to get the preliminary estimates of the primordial 
helium abundance, $Y_p = 0.244\pm0.019$, and the galactic enrichment ratio, $\Delta Y / \Delta Z = 1.226\pm0.849$. Although the current estimates 
have large errorbars due to the limited sample size, this method holds great promises to determine these parameters precisely in the era of upcoming 
space missions.
\end{abstract}


\section{Introduction}
\label{intro}
The observed oscillation frequencies of the stars from the {\it Convection, Rotation and planetary Transits} \citep[{\it CoRoT};][]{bagl09} and 
{\it Kepler} \citep{boru09} space missions are expected to contain enormous amount of information about the stellar interiors 
\citep[see e.g.][]{ulri86,ulri88,jcd88b,jcd02,aert10}, and have been recently used successfully to study them \citep[e.g.][just to name a 
few]{math12,metc12,metc14,silv13,silv15,lebr14,verm14a,buld16a}. Unfortunately, the current methods of fitting the stellar oscillation frequencies 
\citep[see e.g.][]{metc09,grub12,silv15,verm16,bell16} do not exploit the full diagnostic potential of the precisely measured oscillation 
frequencies due to our poor understanding of the near-surface layers -- which introduces a frequency-dependent systematic offset in the model 
frequencies, known as ``surface effect" \citep[see][]{jcd88c,kjel08,ball14} -- and consequently some of the stellar parameters are not as tightly 
constrained as they otherwise would be \citep[see e.g.][]{nsam18}. 

The initial helium abundance, $Y_i$, is one of the least constrained stellar parameter, and is either determined by assuming a galactic enrichment 
ratio, $\Delta{Y}/\Delta{Z}$, from the galactic chemical evolution or by adjusting it to fit the spectroscopic and seismic data. There is currently 
no consensus on the value of $\Delta{Y}/\Delta{Z}$, and different measurements provide significantly different values, typically in the range 
$0.8-3.0$ \citep[see e.g.][]{riba00a,lebr01,peim02,jime03,bals06,casa07}. Consequently very different relationships between $Y_i$ and initial metal 
mass fraction, $Z_i$, have been used in the recent past to construct the evolutionary tracks and isochrones: for example, \citet{dema04} uses 
$Y_i = 0.23 + 2 Z_i$, \citet{dott08} $Y_i = 0.245 + 1.54 Z_i$, \citet{mari08} $Y_i = 0.23 + 2.23 Z_i$, \citet{choi16} $Y_i = 0.249 + 1.5 Z_i$ and 
\citet{hida18} uses $Y_i = 0.247 + 1.31 Z_i$. While determining $Y_i$ by fitting the observed data seems more plausible, it is known to provide 
biased estimates \citep{metc15}, even sub-primordial values for some stars \citep[see e.g.][]{math12,metc14}. Nsamba et al. (in prep.) presents a 
comprehensive exploration of the systematic uncertainties on stellar parameters arising from the treatment of initial helium abundance in the stellar
model grids.

The first and second ionizations of the helium atom leads to a peak in the first adiabatic index, $\Gamma_1$, between the two ionization zones, which 
introduces an oscillatory signature in the oscillation frequency, $\nu$ \citep[see][]{goug88a,voro88,goug90a,broo14,verm14b}. The recent detections 
of this signature in the oscillation frequencies of several main-sequence \citep[see e.g.][]{mazu14,verm17} and red-giant \citep{cors15} stars open 
the door for a semi-direct method to estimate the surface helium abundance, $Y_s$. The amplitude of the oscillatory signature depends on the amount 
of the helium present in its ionization zones. The theoretical investigations in the past showed the possibility of determining $Y_s$ using the 
amplitude of the helium glitch signature \citep{basu04,houd04,mont05}. More recently, \citet{verm14a} used this method for the first time to 
determine the $Y_s$ of a binary system, 16 Cyg A \& B (KIC 12069424 and 12069449), observed by the {\it Kepler} satellite. Our estimates of $Y_s$ 
for 16 Cyg A \& B were used to constrain their internal structure and derive accurately the global properties 
\citep[see e.g.][]{buld15,buld16a,buld16b}.

The amplitudes of the glitch signatures are very small, and require a large set of precisely determined oscillation frequencies to measure them 
reliably. Recently, \citet{lund17} identified a set of main-sequence stars with the highest signal-to-noise (S/N) data from the {\it Kepler} space 
mission, and dubbed as the ``{\it Kepler} asteroseismic LEGACY sample". They computed the oscillation frequencies for each star in the sample, 
which were then used together with the spectroscopic data by a team of modellers to accurately characterize them \citep{silv17}. \citet{verm17} 
performed the glitch analysis for all the stars in the LEGACY sample to estimate the locations of the helium ionization zone and the base of the 
envelope convection zone. In the current work, we extend our earlier analysis on the LEGACY sample to calibrate the observed amplitude of the 
helium signature against the corresponding amplitude in a set of model frequencies of different $Y_s$ to determine the envelope helium abundance 
of the stars. We further infer the initial abundances using the determined surface values and settling predicted by the stellar models, and derive 
the galactic enrichment ratio.

The rest of the paper is organized as follows. We define the average amplitude of the helium signature, and describe briefly the methods to obtain 
it from the oscillation frequencies in Section~\ref{amp}. The input physics of the models used for the calibration and the process to retrieve them 
are outlined in Section~\ref{model}. We present the results in Section~\ref{results}. The conclusions of the paper are summarized in 
Section~\ref{summ}.

\begin{table*}
\centering
\scriptsize
\caption{A summary of the input physics used in the different sets of calibration models.}
\label{tab1}
\begin{tabular}{cccccccc}
\hline\hline
  Model  &  Opacity  &  Equation of state  &  Solar metallicity mixture  &  Nuclear reaction rates  &  Diffusion  &  Overshoot  &  Frequencies\\
\hline
  MESA  &  OP05+Ferg05  &  OPAL05  &  GS98  &  NACRE  &  Thoul94  &  Herwig00  &  ADIPLS\\ 
  GARSTEC  &  OPAL96+Ferg05  &  OPAL05  &  GS98  &  NACRE  &  --  &  Freytag96  &  ADIPLS\\
  YREC  &  OPAL96+Ferg05  &  OPAL05  &  GS98  &  Adel98  &  Thoul94  &  Maeder75  &  Antia94\\
\hline
\end{tabular}
\flushleft{References: OP05 \citep{badn05,seat05}, OPAL96 \citep{igle96}, Ferg05 \citep{ferg05}, OPAL05 \citep[2005 version of][]{roge02}, 
GS98 \citep{gs98}, NACRE \citep{angu99}, Adel98 \citep{adel98}, Thoul94 \citep{thou94}, Herwig00 \citep{herw00}, Freytag96 \citep{frey96}, 
Maeder75 \citep{maed75}, ADIPLS \citep{jcd08a}, Antia94 \citep{anti94}}
\end{table*}

\section{Amplitude of the helium signature}
\label{amp}
The functional form of the helium glitch signature can be obtained using the variational principle \citep{chan64} by assuming a Gaussian profile 
for the $\Gamma_1$-peak between the two helium ionization zones \citep{goug02,houd07}, and can be written as,
\begin{equation}
\delta\nu_{\rm He} = A_{\rm He} \nu e^{-8\pi^2\Delta_{\rm He}^2\nu^2} \sin(4\pi\tau_{\rm He}\nu + \psi_{\rm He}),
\label{helium}
\end{equation}
where the parameter $A_{\rm He}$ is related to the area under the $\Gamma_1$-peak, $\Delta_{\rm He}$ is related to the width, $\tau_{\rm He}$ 
is the acoustic depth of the peak and the parameter $\psi_{\rm He}$ is a phase constant.

We can see from Eq.~\ref{helium} that the amplitude depends on the oscillation frequency. Different authors in the literature have used different 
measures of the amplitude for the calibration. For instance, \citet{migl03} suggested to calibrate the area of the $\Gamma_1$ depression 
in the second helium ionization zone, while \citet{houd07} used the height of the depression for the calibration. In the current interpretation 
of the helium glitch, they would correspond to the area and height of the peak between the two stages of the helium ionization, and are related to 
the parameters $A_{\rm He}$ and $A_{\rm He}/\Delta_{\rm He}$, respectively. \citet{mont05} proposed to use the amplitude at a reference frequency 
($A_{\nu_0} = A_{\rm He} \nu_0 e^{-8\pi^2\Delta_{\rm He}^2\nu_0^2}$, where $\nu_0$ is the reference frequency), whereas \citet{basu04} used the 
amplitude averaged over the observed frequency range for the calibration \citep[see also][]{verm14a}. In this work, we used the average amplitude,
\begin{eqnarray}
\langle A_{\nu} \rangle &=& \frac{\int_{\nu_1}^{\nu_2} A_{\rm He} \nu e^{-8\pi^2\Delta_{\rm He}^2\nu^2} d\nu}{\int_{\nu_1}^{\nu_2} d\nu}\nonumber\\
&=& \frac{A_{\rm He} [e^{-8\pi^2\Delta_{\rm He}^2\nu_1^2} - e^{-8\pi^2\Delta_{\rm He}^2\nu_2^2}]}{16\pi^2\Delta_{\rm He}^2[\nu_2 - \nu_1]},
\label{amplitude}
\end{eqnarray}
where $\nu_1$ and $\nu_2$ are the smallest and largest observed frequencies used in the fit (the same $\nu_1$ and $\nu_2$ are consistently used to 
calculate model $\langle A_{\nu} \rangle$). We shall discuss the advantage of using $\langle A_{\nu} \rangle$ for the calibration in the 
Appendix~\ref{adv}. 

The extraction of the low-amplitude glitch signatures from the oscillation frequency involves non-linear optimization. To test the convergence of
the fitting method, and the robustness of the final estimate of $Y_s$, we extracted the helium glitch signature using two different methods. The 
first (Method A) fitted the oscillation frequencies directly \citep[see][]{mont94,mont98,mont00,verm14a}, while the second (Method B) fitted the 
second differences of the oscillation frequencies with respect to the radial order, $\delta^2\nu_{n,l} = \nu_{n-1,l} - 2\nu_{n,l} + \nu_{n+1,l}$, 
where $n$ and $l$ are the radial order and harmonic degree, respectively \citep[see][]{goug90a,basu94,basu04,verm14a}. The details of the methods 
used in the current work had been described in \citet{verm14a}. Here, we only outline them for the reader and highlight the minor differences from
the previous works.

\subsection{Fitting the frequencies directly (Method A)}
We modelled the smooth component of the oscillation frequencies corresponding to the smooth structure of the star using a $l$-dependent fourth 
degree polynomial in $n$,
\begin{equation}
\nu_{\rm smooth}(n,l) = \sum_{k = 0}^4 b_k(l) n^k.
\end{equation}
The degree dependence of the polynomial coefficients, $b_k(l)$, simply means that they are different for the different $l$, and need to be 
determined. We determined these coefficients together with the parameters associated with the glitch signatures -- including $A_{\rm He}$ and 
$\Delta_{\rm He}$ required for the average amplitude -- by fitting the oscillation frequencies to the function,
\begin{equation}
f(n,l) = \nu_{\rm smooth} + \delta\nu_{\rm He} + \delta\nu_{\rm CZ},
\label{fitting_function_a}
\end{equation}
where $\delta\nu_{\rm CZ} = \frac{A_{\rm CZ}}{\nu^2} \sin(4\pi\tau_{\rm CZ}\nu + \psi_{\rm CZ})$ is the glitch signature arising from the base of 
the envelope convection zone. The parameter $A_{\rm CZ}$ is related to the jump in the second derivative of the sound speed at the base of the 
convection zone, $\tau_{\rm CZ}$ is the acoustic depth of the convection zone base and $\psi_{\rm CZ}$ is a phase constant. 

The fitting was accomplished by minimizing a cost function,
\begin{equation}
\chi_{\rm A}^2 = \sum_{n, l} \left[\frac{\nu_{n,l} - f(n,l)}{\sigma_{n,l}}\right]^2 + \lambda_{\rm A}^2 \sum_{n,l}\left[\frac{d^3\nu_{\rm smooth}}{dn^3}\right]^2,
\end{equation}
where $\sigma_{n,l}$ is the uncertainty on $\nu_{n,l}$ and $\lambda_{\rm A}$ a regularization parameter. Note that the third derivative 
regularization used in this work \citep[also used in][]{verm17} marginally improves the stability of the fit in comparison to the second derivative
used in \citet{verm14a,verm14b}. Previous work has shown that $\lambda_{\rm A} = 7$ is a suitable choice \citep{verm17}. While fitting the models 
we used the same set of modes and weights as for the observations. We propagated the statistical uncertainties on the observed oscillation 
frequencies to the glitch parameters by fitting 1000 realizations of the data (Monte-Carlo simulation). The model frequencies do not have statistical
uncertainties, however they are known to be affected by the systematic uncertainties because of the uncertainties in the stellar and modal physics. 
A combination of these uncertainties show up as surface effect. We shall quantify the impact of the surface term on the glitch parameters in the 
Appendix~\ref{surface}, and show that its effect is small.

\subsection{Fitting the second differences (Method B)}
Motivated from the asymptotic theory of the stellar oscillations \citep{tass80} we assumed that the smooth component of the second differences 
is independent of $l$, and modelled it with a quadratic function of the frequency, $\delta^2\nu_{\rm smooth} = a_0 + a_1 \nu + a_2 \nu^2$. In this 
method, the glitch parameters were determined by fitting the second differences of the oscillation frequencies to the function,
\begin{eqnarray}
g(n,l) &=& \delta^2\nu_{\rm smooth} + a_{\rm He} \nu e^{-8\pi^2\Delta_{\rm He}^2\nu^2} \sin(4\pi\tau_{\rm He}\nu + \phi_{\rm He})\nonumber\\
&+&\frac{a_{\rm CZ}}{\nu^2} \sin(4\pi\tau_{\rm CZ}\nu + \phi_{\rm CZ}),
\label{fitting_function_b}
\end{eqnarray}
where the second and third term represent the helium and convection zone signature, respectively. Note that the amplitudes, $a_{\rm He}$ and 
$a_{\rm CZ}$, and the phases, $\phi_{\rm He}$ and $\phi_{\rm CZ}$, of the signatures in the second differences are different from the corresponding 
amplitudes and phases of the signatures in the oscillation frequencies (see Eq.~\ref{fitting_function_a}). The amplitude of the helium signature in 
the second differences is related to the amplitude in the frequencies, 
$a_{\rm He} \approx 4 A_{\rm He} \sin^2(2\pi\tau_{\rm He}\langle \Delta\nu \rangle)$, where $\langle \Delta\nu \rangle$ is the average large 
frequency separation \citep[see][]{basu94}. 

We determined $a_0$, $a_1$, $a_2$ and the glitch parameters by minimizing a cost function,
\begin{equation}
\chi_{\rm B}^2 = \mathbfit{x}^T \mathbfss{C}^{-1} \mathbfit{x} + \lambda_{\rm B}^2 \sum_{n,l}\left[\frac{d\delta^2\nu_{\rm smooth}}{d\nu}\right]^2,
\end{equation}
where $\mathbfit{x}$ is a vector containing the differences between the observed and model second differences, $\mathbfss{C}$ the analytic covariance 
matrix for the second differences and $\lambda_{\rm B}$ is a regularization parameter. In this method, the choice of the degree of polynomial for
the smooth component and the order of derivative in the regularization term are both inspired by Method A. Since we are fitting the second 
difference of the oscillation frequency, the degree of polynomial and the order of derivative were both reduced by two to second degree polynomial
and first order derivative. We used the data for 16 Cyg A \& B and followed the procedure described in \citet{verm14a} to find an optimal value of 
$\lambda_{\rm B} = 1000$. Again, the same set of modes and weights were used while fitting the model frequencies. We propagated the observational 
uncertainties on the oscillation frequencies to the glitch parameters using the Monte-Carlo simulation.

\begin{table*}
\centering
\scriptsize
\caption{The initial parameter space covered by the MESA and YREC (non-diffusion set) for each star. The solar calibrated mixing-length for the 
MESA and YREC with the input physics described in Section~\ref{model} are 1.91 (1.84 when excluding the atomic diffusion) and 1.70, respectively. 
The missing range for $f_{\rm OV}$ for some stars means that the overshoot was not included. For the YREC, $\alpha_{\rm OV}$ was fixed at 
0.2$H_p$, where $H_p$ is the pressure scale height.}
\label{tab2}
\begin{tabular}{ccccccccccc}
\hline\hline
  KIC  &  \multicolumn{5}{c}{MESA}  &  &  \multicolumn{4}{c}{YREC}\\
\cline{2-6} \cline{8-11}\\
       &  $M$ (M$_\odot$)  &  $Y_i$  &  $[{\rm Fe}/{\rm H}]_i$  &  $\alpha_{\rm MLT}$  &  $f_{\rm OV}$  &  &  $M$ (M$_\odot$)  &  $Y_i$  &  
$[{\rm Fe}/{\rm H}]_i$  &  $\alpha_{\rm MLT}$\\
\hline
  1435467  &  1.25--1.55  &  0.22--0.32  &  $-0.10$ -- $+0.10$  &  1.5--2.0  &  0.00--0.03  &  &  1.30--1.55  &  0.16--0.34  &  $-0.10$ -- $+0.25$  &  1.60--2.10\\ 
  2837475  &  1.30--1.60  &  0.22--0.32  &  $-0.10$ -- $+0.10$  &  1.5--2.0  &  0.00--0.03  &  &  1.40--1.90  &  0.05--0.32  &  $-0.08$ -- $+0.18$  &  1.65--1.90\\
  3427720  &  1.00--1.30  &  0.22--0.32  &  $-0.05$ -- $+0.15$  &  1.5--2.0  &  0.00--0.03  &  &  1.05--1.40  &  0.10--0.28  &  $-0.15$ -- $+0.05$  &  1.50--2.10\\
  3456181  &  1.35--1.65  &  0.22--0.32  &  $-0.25$ -- $-0.05$  &  1.5--2.0  &  0.00--0.03  &  &  1.40--1.80  &  0.10--0.28  &  $-0.20$ -- $+0.00$  &  1.60--2.10\\
  3632418  &  1.05--1.35  &  0.22--0.32  &  $-0.10$ -- $+0.10$  &  1.5--2.0  &  0.00--0.05  &  &  1.30--1.55  &  0.12--0.28  &  $-0.20$ -- $+0.10$  &  1.40--1.95\\ 
  3656476  &  1.00--1.30  &  0.24--0.34  &  $+0.30$ -- $+0.50$  &  1.5--2.0  &  0.00--0.03  &  &  1.02--1.20  &  0.20--0.30  &  $+0.05$ -- $+0.35$  &  1.45--1.90\\
  3735871  &  1.00--1.30  &  0.22--0.32  &  $-0.10$ -- $+0.10$  &  1.5--2.0  &  0.00--0.03  &  &  1.05--1.35  &  0.12--0.30  &  $-0.15$ -- $+0.05$  &  1.50--2.10\\
  4914923  &  1.00--1.30  &  0.22--0.32  &  $+0.15$ -- $+0.35$  &  1.5--2.0  &  0.00--0.03  &  &  0.96--1.22  &  0.16--0.32  &  $-0.10$ -- $+0.30$  &  1.20--1.90\\
  5184732  &  1.05--1.35  &  0.24--0.34  &  $+0.35$ -- $+0.55$  &  1.5--2.0  &  0.00--0.03  &  &  1.15--1.50  &  0.15--0.34  &  $+0.15$ -- $+0.45$  &  1.45--1.90\\ 
  5773345  &  1.30--1.60  &  0.22--0.32  &  $+0.10$ -- $+0.30$  &  1.5--2.0  &  0.00--0.03  &  &  1.45--1.80  &  0.12--0.32  &  $+0.10$ -- $+0.40$  &  1.40--2.10\\
  5950854  &  0.80--1.10  &  0.22--0.32  &  $-0.20$ -- $+0.00$  &  1.5--2.0  &    \dots     &  &  0.95--1.20  &  0.08--0.26  &  $-0.40$ -- $-0.05$  &  1.40--2.10\\
  6106415  &  0.95--1.25  &  0.22--0.32  &  $-0.10$ -- $+0.10$  &  1.5--2.0  &    \dots     &  &  1.06--1.24  &  0.18--0.27  &  $-0.20$ -- $+0.05$  &  1.55--1.90\\
  6116048  &  0.95--1.25  &  0.20--0.30  &  $-0.20$ -- $+0.00$  &  1.5--2.0  &    \dots     &  &  1.00--1.20  &  0.13--0.26  &  $-0.40$ -- $-0.05$  &  1.45--1.90\\
  6225718  &  1.00--1.30  &  0.22--0.32  &  $-0.15$ -- $+0.05$  &  1.5--2.0  &  0.00--0.05  &  &  1.15--1.45  &  0.14--0.28  &  $-0.20$ -- $+0.05$  &  1.60--2.10\\
  6508366  &  1.35--1.65  &  0.22--0.32  &  $-0.15$ -- $+0.05$  &  1.5--2.0  &  0.00--0.03  &  &  1.45--1.80  &  0.10--0.30  &  $-0.10$ -- $+0.10$  &  1.65--1.90\\
  6603624  &  0.90--1.20  &  0.22--0.32  &  $+0.25$ -- $+0.45$  &  1.5--2.0  &    \dots     &  &  0.95--1.20  &  0.16--0.30  &  $-0.00$ -- $+0.35$  &  1.40--2.10\\
  6679371  &  1.35--1.65  &  0.22--0.32  &  $-0.10$ -- $+0.10$  &  1.5--2.0  &  0.00--0.03  &  &  1.40--1.65  &  0.24--0.34  &  $-0.10$ -- $+0.20$  &  1.45--1.90\\
  6933899  &  1.00--1.30  &  0.22--0.32  &  $+0.05$ -- $+0.25$  &  1.5--2.0  &  0.00--0.03  &  &  1.05--1.30  &  0.14--0.28  &  $-0.15$ -- $+0.10$  &  1.25--1.50\\
  7103006  &  1.15--1.35  &  0.22--0.32  &  $+0.05$ -- $+0.25$  &  1.5--2.0  &  0.00--0.03  &  &  1.30--1.75  &  0.10--0.35  &  $-0.10$ -- $+0.20$  &  1.65--1.90\\
  7106245  &  0.70--1.00  &  0.22--0.32  &  $-0.85$ -- $-0.55$  &  1.5--2.0  &    \dots     &  &  0.90--1.15  &  0.06--0.26  &  $-0.90$ -- $-0.45$  &  1.20--2.00\\
  7206837  &  1.25--1.55  &  0.22--0.32  &  $+0.05$ -- $+0.25$  &  1.5--2.0  &  0.00--0.05  &  &  1.25--1.65  &  0.10--0.35  &  $-0.00$ -- $+0.25$  &  1.45--1.90\\
  7296438  &  1.00--1.30  &  0.22--0.32  &  $+0.20$ -- $+0.40$  &  1.5--2.0  &  0.00--0.03  &  &  1.05--1.30  &  0.16--0.30  &  $+0.05$ -- $+0.35$  &  1.45--1.90\\
  7510397  &  1.15--1.40  &  0.22--0.32  &  $-0.15$ -- $+0.15$  &  1.5--2.0  &  0.00--0.05  &  &  1.30--1.50  &  0.14--0.32  &  $-0.30$ -- $+0.10$  &  1.40--1.90\\
  7680114  &  1.00--1.30  &  0.22--0.32  &  $+0.10$ -- $+0.30$  &  1.5--2.0  &  0.00--0.03  &  &  1.05--1.25  &  0.16--0.28  &  $-0.10$ -- $+0.15$  &  1.40--1.90\\
  7771282  &  1.15--1.35  &  0.22--0.32  &  $+0.00$ -- $+0.20$  &  1.5--2.0  &  0.00--0.03  &  &  1.20--1.55  &  0.16--0.32  &  $-0.10$ -- $+0.10$  &  1.60--1.90\\
  7871531  &  0.70--1.00  &  0.22--0.32  &  $-0.25$ -- $+0.05$  &  1.5--2.0  &    \dots     &  &  0.80--1.05  &  0.08--0.30  &  $-0.35$ -- $-0.10$  &  1.50--2.10\\
  7940546  &  1.20--1.40  &  0.22--0.32  &  $-0.05$ -- $+0.15$  &  1.5--2.0  &  0.00--0.03  &  &  1.35--1.55  &  0.10--0.30  &  $-0.30$ -- $+0.05$  &  1.40--1.90\\
  7970740  &  0.65--0.95  &  0.22--0.32  &  $-0.50$ -- $-0.20$  &  1.5--2.0  &    \dots     &  &  0.70--0.90  &  0.10--0.28  &  $-0.70$ -- $-0.35$  &  1.40--2.10\\
  8006161  &  0.80--1.10  &  0.22--0.32  &  $+0.30$ -- $+0.50$  &  1.5--2.0  &    \dots     &  &  0.90--1.10  &  0.22--0.33  &  $+0.15$ -- $+0.45$  &  1.60--2.10\\
  8150065  &  1.00--1.30  &  0.22--0.32  &  $-0.10$ -- $+0.10$  &  1.5--2.0  &  0.00--0.03  &  &  1.10--1.70  &  0.05--0.30  &  $-0.20$ -- $+0.05$  &  1.40--1.90\\
  8179536  &  1.15--1.35  &  0.22--0.32  &  $+0.00$ -- $+0.20$  &  1.5--2.0  &  0.00--0.03  &  &  1.20--1.70  &  0.05--0.32  &  $-0.10$ -- $+0.10$  &  1.60--2.10\\
  8228742  &  1.05--1.35  &  0.20--0.30  &  $-0.05$ -- $+0.15$  &  1.5--2.0  &  0.00--0.03  &  &  1.30--1.50  &  0.16--0.30  &  $-0.20$ -- $+0.05$  &  1.45--1.90\\
  8379927  &  1.00--1.30  &  0.20--0.30  &  $-0.20$ -- $+0.00$  &  1.5--2.0  &  0.00--0.05  &  &  1.10--1.30  &  0.14--0.30  &  $-0.30$ -- $+0.30$  &  1.40--2.10\\
  8394589  &  0.90--1.20  &  0.22--0.32  &  $-0.30$ -- $-0.10$  &  1.5--2.0  &    \dots     &  &  1.00--1.20  &  0.15--0.28  &  $-0.40$ -- $-0.15$  &  1.45--1.70\\
  8424992  &  0.70--1.00  &  0.22--0.32  &  $-0.10$ -- $+0.10$  &  1.5--2.0  &    \dots     &  &  0.90--1.10  &  0.10--0.30  &  $-0.30$ -- $+0.10$  &  1.60--2.10\\
  8694723  &  0.95--1.25  &  0.20--0.30  &  $-0.45$ -- $-0.15$  &  1.5--2.0  &  0.00--0.03  &  &  1.10--1.30  &  0.18--0.23  &  $-0.50$ -- $-0.20$  &  1.40--1.90\\
  8760414  &  0.70--1.00  &  0.20--0.30  &  $-1.05$ -- $-0.60$  &  1.5--2.0  &    \dots     &  &  0.80--0.91  &  0.14--0.25  &  $-1.05$ -- $-0.55$  &  1.30--1.90\\
  8938364  &  0.80--1.10  &  0.22--0.32  &  $-0.15$ -- $+0.15$  &  1.5--2.0  &    \dots     &  &  0.90--1.10  &  0.20--0.29  &  $-0.30$ -- $+0.10$  &  1.20--1.70\\
  9025370  &  0.75--1.05  &  0.22--0.32  &  $-0.10$ -- $+0.10$  &  1.5--2.0  &    \dots     &  &  0.90--1.25  &  0.05--0.32  &  $-0.20$ -- $+0.20$  &  1.20--1.70\\
  9098294  &  0.85--1.15  &  0.22--0.32  &  $-0.18$ -- $+0.02$  &  1.5--2.0  &    \dots     &  &  0.98--1.12  &  0.14--0.28  &  $-0.30$ -- $+0.05$  &  1.60--1.90\\
  9139151  &  1.00--1.30  &  0.22--0.32  &  $+0.05$ -- $+0.25$  &  1.5--2.0  &  0.00--0.05  &  &  1.10--1.35  &  0.16--0.32  &  $-0.10$ -- $+0.20$  &  1.80--2.30\\
  9139163  &  1.30--1.60  &  0.22--0.32  &  $+0.05$ -- $+0.25$  &  1.5--2.0  &  0.00--0.03  &  &  1.30--1.70  &  0.15--0.35  &  $-0.00$ -- $+0.40$  &  1.60--2.10\\
  9206432  &  1.35--1.65  &  0.22--0.32  &  $+0.05$ -- $+0.25$  &  1.5--2.0  &  0.00--0.03  &  &  1.30--2.00  &  0.10--0.34  &  $+0.10$ -- $+0.35$  &  1.80--2.10\\
  9353712  &  1.25--1.55  &  0.22--0.32  &  $-0.15$ -- $+0.05$  &  1.5--2.0  &  0.00--0.03  &  &  1.45--1.70  &  0.16--0.30  &  $-0.15$ -- $+0.15$  &  1.60--1.90\\
  9410862  &  0.80--1.10  &  0.22--0.32  &  $-0.30$ -- $-0.10$  &  1.5--2.0  &    \dots     &  &  0.95--1.10  &  0.16--0.30  &  $-0.45$ -- $-0.15$  &  1.45--1.90\\
  9414417  &  1.10--1.30  &  0.22--0.32  &  $-0.10$ -- $+0.10$  &  1.5--2.0  &  0.00--0.03  &  &  1.35--1.70  &  0.05--0.30  &  $-0.25$ -- $+0.05$  &  1.65--1.80\\
  9812850  &  1.15--1.35  &  0.22--0.32  &  $-0.05$ -- $+0.15$  &  1.5--2.0  &  0.00--0.03  &  &  1.30--1.70  &  0.10--0.30  &  $-0.20$ -- $+0.10$  &  1.60--2.10\\
  9955598  &  0.70--1.00  &  0.22--0.32  &  $+0.10$ -- $+0.30$  &  1.5--2.0  &    \dots     &  &  0.85--1.05  &  0.16--0.30  &  $-0.10$ -- $+0.25$  &  1.60--2.10\\
  9965715  &  0.80--1.10  &  0.22--0.32  &  $-0.35$ -- $-0.05$  &  1.5--2.0  &    \dots     &  &  1.00--1.40  &  0.00--0.28  &  $-0.55$ -- $-0.10$  &  1.20--1.70\\
 10068307  &  1.20--1.40  &  0.22--0.32  &  $-0.15$ -- $+0.05$  &  1.5--2.0  &  0.00--0.03  &  &  1.40--1.75  &  0.05--0.28  &  $-0.30$ -- $+0.10$  &  1.20--2.00\\
 10079226  &  1.00--1.30  &  0.22--0.32  &  $+0.05$ -- $+0.25$  &  1.5--2.0  &  0.00--0.03  &  &  1.05--1.30  &  0.16--0.32  &  $-0.00$ -- $+0.20$  &  1.40--1.90\\
 10162436  &  1.05--1.40  &  0.22--0.32  &  $-0.05$ -- $+0.15$  &  1.5--2.0  &  0.00--0.05  &  &  1.35--1.60  &  0.14--0.30  &  $-0.20$ -- $+0.10$  &  1.40--1.90\\
 10454113  &  1.00--1.30  &  0.22--0.32  &  $-0.05$ -- $+0.15$  &  1.5--2.0  &  0.00--0.05  &  &  1.10--1.40  &  0.12--0.30  &  $-0.10$ -- $+0.10$  &  1.40--1.70\\
 10516096  &  1.00--1.30  &  0.22--0.32  &  $-0.10$ -- $+0.10$  &  1.5--2.0  &  0.00--0.03  &  &  1.05--1.25  &  0.15--0.26  &  $-0.25$ -- $+0.00$  &  1.40--1.90\\
 10644253  &  1.00--1.30  &  0.22--0.32  &  $+0.10$ -- $+0.30$  &  1.5--2.0  &  0.00--0.03  &  &  1.05--1.35  &  0.16--0.30  &  $-0.05$ -- $+0.20$  &  1.45--1.90\\
 10730618  &  1.15--1.35  &  0.22--0.32  &  $-0.10$ -- $+0.10$  &  1.5--2.0  &  0.00--0.03  &  &  1.20--1.60  &  0.05--0.33  &  $-0.20$ -- $+0.25$  &  1.20--2.10\\
 10963065  &  0.90--1.20  &  0.22--0.32  &  $-0.15$ -- $+0.05$  &  1.5--2.0  &    \dots     &  &  1.00--1.20  &  0.16--0.28  &  $-0.35$ -- $-0.10$  &  1.45--1.75\\
 11081729  &  1.20--1.40  &  0.22--0.32  &  $+0.15$ -- $+0.35$  &  1.5--2.0  &  0.00--0.03  &  &  1.20--1.60  &  0.12--0.34  &  $-0.00$ -- $+0.25$  &  1.80--2.20\\
 11253226  &  1.30--1.60  &  0.22--0.32  &  $-0.20$ -- $+0.00$  &  1.5--2.0  &  0.00--0.03  &  &  1.30--1.90  &  0.00--0.33  &  $-0.15$ -- $+0.15$  &  1.65--2.10\\
 11772920  &  0.70--1.00  &  0.22--0.32  &  $-0.10$ -- $+0.10$  &  1.5--2.0  &    \dots     &  &  0.80--1.05  &  0.00--0.30  &  $-0.30$ -- $+0.20$  &  1.40--2.10\\
 12009504  &  1.00--1.30  &  0.22--0.32  &  $-0.05$ -- $+0.15$  &  1.5--2.0  &  0.00--0.03  &  &  1.15--1.45  &  0.08--0.32  &  $-0.25$ -- $+0.10$  &  1.40--2.10\\
 12069127  &  1.35--1.65  &  0.22--0.32  &  $+0.00$ -- $+0.20$  &  1.5--2.0  &  0.00--0.03  &  &  1.45--1.85  &  0.14--0.32  &  $-0.00$ -- $+0.20$  &  1.60--1.90\\
 12069424  &  0.85--1.15  &  0.22--0.32  &  $+0.10$ -- $+0.30$  &  1.5--2.0  &    \dots     &  &  1.00--1.20  &  0.20--0.27  &  $+0.02$ -- $+0.16$  &  1.55--1.90\\
 12069449  &  0.85--1.15  &  0.22--0.32  &  $+0.05$ -- $+0.25$  &  1.5--2.0  &    \dots     &  &  0.95--1.15  &  0.19--0.31  &  $-0.00$ -- $+0.20$  &  1.50--2.10\\
 12258514  &  1.00--1.30  &  0.22--0.32  &  $+0.05$ -- $+0.25$  &  1.5--2.0  &  0.00--0.05  &  &  1.20--1.45  &  0.10--0.28  &  $-0.10$ -- $+0.15$  &  1.40--2.00\\
 12317678  &  1.10--1.30  &  0.22--0.32  &  $-0.15$ -- $+0.05$  &  1.5--2.0  &  0.00--0.03  &  &  1.20--1.70  &  0.05--0.32  &  $-0.40$ -- $-0.10$  &  1.65--2.10\\
\hline
\end{tabular}
\end{table*}

\section{Calibration models}
\label{model}
We used four different sets of the calibration models: one with the Modules for Experiments in Stellar Astrophysics \citep[MESA;][]{paxt11,paxt13}, 
one with the Garching Stellar Evolution Code \citep[GARSTEC;][]{weis08} and two with the Yale Rotating Stellar Evolution Code 
\citep[YREC;][]{dema08}. This was done to quantify the systematic uncertainties on inferred $Y_s$ associated with the choice of the input physics. 
A summary of the input physics used in these codes is provided in Table~\ref{tab1}. The second set of the YREC models use the same input physics 
as the one in the table, except that it does not include atomic diffusion. The details of the input physics and the procedure to get the 
calibration models are described below for all the sets.

\subsection{MESA models}
\label{mesa}
We used MESA with Opacity Project (OP) high-temperature opacities \citep{badn05,seat05} supplemented with low-temperature opacities of
\citet{ferg05}. The metallicity mixture from \citet{gs98} was used. We used OPAL equation of state \citep{roge02}. The reaction rates were from 
NACRE \citep{angu99} for all reactions except $^{14}{\rm N}(p,\gamma)^{15}{\rm O}$ and $^{12}{\rm C}(\alpha,\gamma)^{16}{\rm O}$, for which updated 
reaction rates from \citet{imbr05} and \citet{kunz02} were used. Since the inclusion of atomic diffusion leads to complete depletion of the surface 
helium and heavy elements for solar metallicity stellar models of masses approximately greater than 1.4 M$_\odot$ \citep[see e.g.][]{more02}, we 
included this process only for the stars of masses less than 1.35 M$_\odot$ using the prescription of \citet{thou94}. It is now well known that 
the stars with masses approximately greater than 1.1 M$_\odot$ have finite convective core overshoot \citep[see e.g.][]{dehe10,de10,silv13}, 
though the value of the overshoot parameter remains uncertain, and may depend on the mass of the star \citep[see e.g.][]{riba00b,clar16}. Hence 
we used an exponential overshoot \citep{herw00} with variable overshoot parameter, $f_{\rm OV}$, for the stars with masses greater than 1.10 
M$_\odot$. The adiabatic oscillation frequencies were calculated using the Adiabatic Pulsation code \citep[ADIPLS;][]{jcd08a}.

In this approach of constructing the calibration models, the choice of input physics depends on the mass of the star at hand. We estimate the
mass using the asteroseismic scaling relations \citep{kjel95}, and use the input physics based on that. The detailed asteroseismic modelling of the 
star eventually provides more accurate value of the mass. We restart the process if the new mass suggests that the input physics needs to be changed.
This is rare but may happen if the mass of the star is close to either 1.10 or 1.35 M$_\odot$.

We computed the tracks for each star with different mass $M$, initial helium abundance $Y_i$, initial metallicity $[{\rm Fe}/{\rm H}]_i$, 
mixing-length $\alpha_{\rm MLT}$ and overshoot parameter $f_{\rm OV}$. We generated 1000--2000 mesh points randomly with uniform distribution 
in the parameter space listed in Table~\ref{tab2}. The initial parameter ranges listed in the table were iteratively modified so that the 
best-fitting model did not fall at the edge of the parameter space. 

We evolved the initial pre-main-sequence model corresponding to every mesh point until the track entered in a box formed by the $4\sigma$ 
uncertainties in the observed effective temperature $T_{\rm eff}$, surface metallicity $[{\rm Fe}/{\rm H}]_s$ and the average large frequency 
separation $\langle\Delta\nu\rangle$. Afterwards, we fitted the surface corrected model frequencies \citep{kjel08} on the fly to the observed ones 
to break the degeneracy inside the box, and accept the best-fitting model as a representative model of the concerned star. Repeating the above 
process for all the mesh points, we got a set of approximately 1000--2000 representative models for each star.

The choice of the surface correction scheme is not important in this particular exercise because, for a given initial condition ($M$, $Y_i$, 
$[{\rm Fe}/{\rm H}]_i$, $\alpha_{\rm MLT}$ and $f_{\rm OV}$), the age can be determined very precisely by fitting the observed oscillation 
frequencies irrespective of the surface correction used. The large uncertainties in the inferred ages of the stars are the result of the 
uncertainties in the initial conditions, particularly in $M$ and $f_{\rm OV}$. Having said that, the YREC models use a different surface 
correction \citep[][see Section~\ref{yrec}]{ball14}, and hence its impact can be estimated by comparing the $Y_s$ from the MESA/GARSTEC and YREC.

\subsection{GARSTEC models}
\label{garstec}
The code was used with OPAL high-temperature opacities \citep{igle96} supplemented with low-temperature opacities of \citet{ferg05}. We used 
solar metallicity mixture from \citet{gs98}. The OPAL equation of state was used \citep{roge02}. All the reaction rates were from NACRE except 
$^{14}{\rm N}(p,\gamma)^{15}{\rm O}$ and $^{12}{\rm C}(\alpha,\gamma)^{16}{\rm O}$ for which the rates from \citet{form04} and \citet{hamm05} were 
used. We did not include atomic diffusion in this case intentionally to formally incorporate the associated systematic uncertainty on the 
quoted values of $Y_s$. The overshoot was included for all the stars following the prescription of \citet{frey96} with variable overshoot parameter,
$f_{\rm OV}$. Note that the expression for the overshoot diffusion coefficient is same for both \citet{frey96} and \citet{herw00}, and hence the 
same associated overshoot parameter. Aarhus adiabatic pulsation package (ADIPLS) was used for the frequency calculations \citep{jcd08a}.

We generated 6000 tracks in the parameter space $M \in [0.7, 1.8] M_\odot$, $Y_i \in [0.15, 0.40]$, $[{\rm Fe}/{\rm H}]_i \in [-1.2, 0.6]$
dex, $\alpha_{\rm MLT} \in [1.5, 2.2]$ and $f_{\rm OV} \in [0.00, 0.03]$. The above space was again populated randomly from uniform distribution, 
but this time using quasi-random number generator \citep{sobo67} instead of pseudo-random number. We stored every third model during the evolution, 
resulting several hundred main-sequence models per track. Note that this way of generating a grid is slightly different from the ones generated 
using MESA, for which the initial parameter space is limited to a target star (particularly the mass and initial metallicity) and only one model -- 
that fits the observed frequencies the best -- is stored per track. The increased parameter space and storing several hundred to a thousand 
models per track for GARSTEC increase the grid size considerably. The advantage of one such giant grid is that we can reuse it for the newly 
observed stars unlike the grid constructed using the MESA. The disadvantage is that we need much larger disk space (current grid needs 4TBs of disk 
space to store the local properties of the models needed for the frequency computation). Furthermore, the models on a track have poorer temporal 
resolution.

We obtained models for the individual stars from the above tracks following the same procedure as for the MESA, i.e. we scanned all the models 
in a track and located the age range where $T_{\rm eff}$, $[{\rm Fe}/{\rm H}]_s$ and $\langle\Delta\nu\rangle$ were all within 4$\sigma$ of the 
observed values, subsequently we fitted the surface corrected model frequencies \citep{kjel08} to get a representative model for the star. 
Repeating this for all the 6000 tracks, we obtain several tens to few hundred models per star.

\subsection{YREC models}
\label{yrec}
We used YREC with OPAL high-temperature opacities \citep{igle96} supplemented with low-temperature opacities of \citet{ferg05}. The
metallicity mixture from \citet{gs98} was used. We used the 2005 version of the OPAL equation of state \citep{roge02}. All the nuclear reaction
rates were from \citet{adel98} except for the $^{14}$N($p$,$\gamma$)$^{15}O$ reaction, for which we used the updated rates of \citet{form04}. A 
step overshoot \citep{maed75} of $\alpha_{\rm OV} = 0.2H_p$, where $H_p$ is the pressure scale height, was used for all the stars. The YREC models 
were used to isolate the impact of the atomic diffusion on $Y_s$ determination by using the models with and without diffusion. The adiabatic 
oscillation frequencies were calculated using a code described in \citet{anti94}.

The Yale Monte-Carlo Method \citep[YMCM;][]{silv15} was used to get a set of representative models for each star. We started with using the 
average large frequency separation and frequency of maximum power along with the spectroscopically determined effective temperature to get an 
estimate of the mass and radius of the star using the Yale Birmingham Grid-Based modelling pipeline \citep{basu10,gai11}. Since each of the 
observables has the associated observational error, we created thousands of realizations of $M$, $R$, $T_{\rm eff}$, and $[{\rm Fe}/{\rm H}]_s$. For 
each realization, we used the YREC in an iterative mode to obtain a model of given $M$ and $[{\rm Fe}/{\rm H}]_s$ that had the required $R$ and 
$T_{\rm eff}$. This was done in two different ways: in the first approach, we kept $\alpha_{\rm MLT}$ fixed at different values and iterated over 
$Y_i$ to get the model; and in the second approach, we kept $Y_i$ fixed at different values and varied $\alpha_{\rm MLT}$ to get the required 
model. Finally, a set of representative models of the star was obtained based on the merit function, 
\begin{equation}
\chi^2_{\rm total} = \chi^2_{\nu} + \chi^2_{\rm ratios} + \chi^2_{\rm Teff} + \chi^2_{[{\rm Fe}/{\rm H}]_s},
\label{chi2total}
\end{equation}
where $\chi^2_\nu$ and $\chi^2_{\rm ratios}$ are the reduced chi-squares for the oscillation frequencies and frequency ratios, $r_{02}$ and $r_{01}$ 
\citep[for the definition of ratios, see][]{roxb03}. The $\chi^2_\nu$ was computed using the surface corrected model frequencies \citep[two term
correction of][]{ball14}. Since ratios are independent of the surface term and are strongly correlated, the $\chi^2_{\rm ratios}$ was calculated 
using the uncorrected model frequencies and using full error covariance matrix.

In this case, the ranges in $M$, $Y_i$, $[{\rm Fe}/{\rm H}]_i$ and $\alpha_{\rm MLT}$ are result of the above process, and are listed in 
Table~\ref{tab2} for only the non-diffusion set.

\section{Results}
\label{results}
We fitted the signatures of the acoustic glitches in all the sets of uncorrected model frequencies (MESA, GARSTEC and YREC) using the both fitting 
methods (A and B). The fits were subsequently used to compute the average amplitude using Eq.~\ref{amplitude}. 

\begin{figure*}
\includegraphics[width=\textwidth]{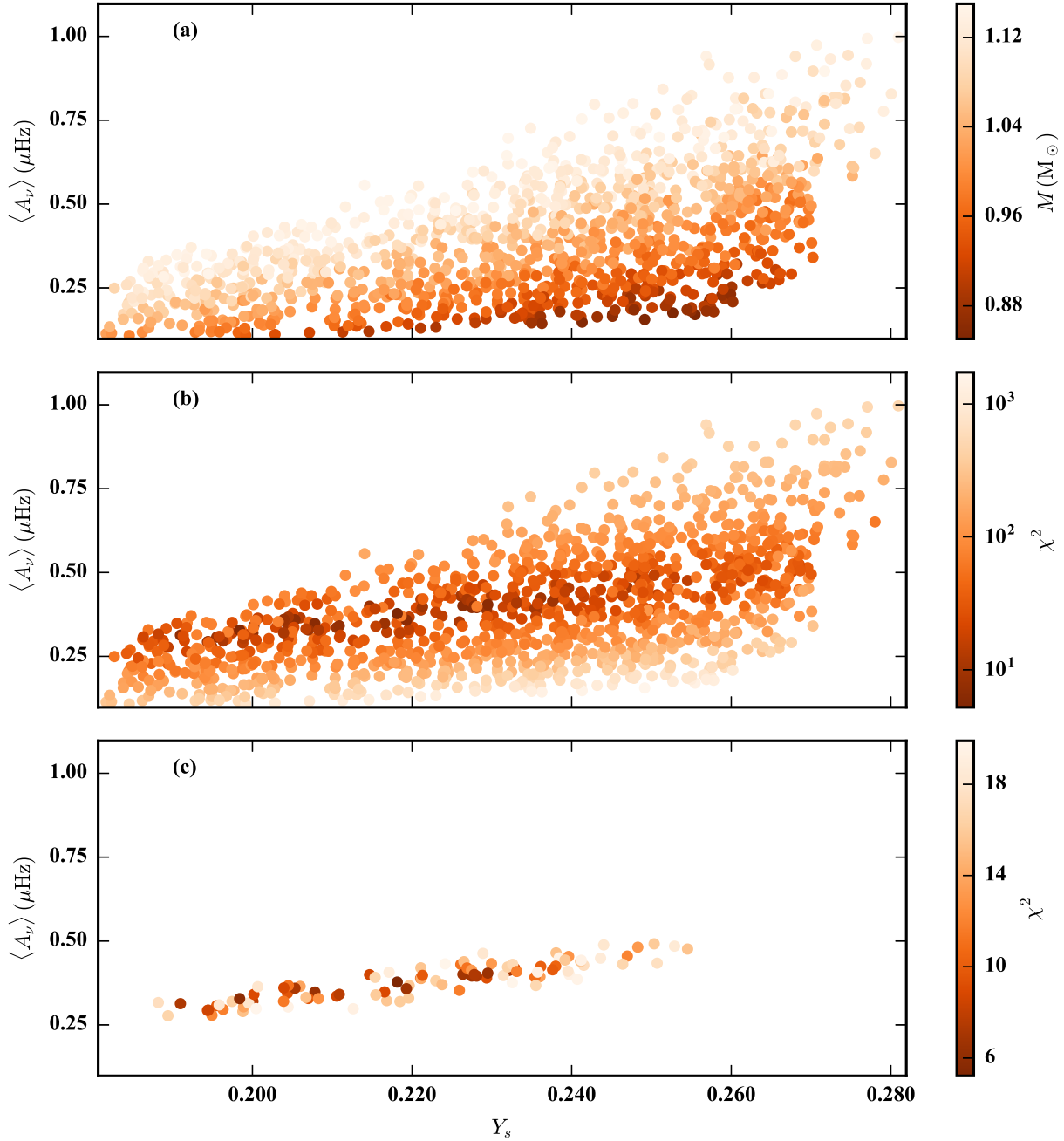}
\caption{Average amplitude of the helium glitch signature obtained using Method A for the models of 16 Cyg A as a function of the envelope helium 
abundance. The points in the panels (a--b) represent all the models obtained using MESA. The points in the panel (c) represent the models with 
$\chi^2$ (as defined in Eq.~\ref{chi2}) less than 20. The colour in the panel (a) shows the mass of the models, while in the panels (b--c) it 
represents their $\chi^2$ value.}
\label{fig1}
\end{figure*}

\subsection{Final set of models used in the calibration}
We filter all the sets of models from Section~\ref{model} homogeneously using the frequency ratios. Note that this does not affect the sets of YREC 
models much, because Eq.~\ref{chi2total} already used the ratios. The process of filtering is illustrated in Figure~\ref{fig1}. We can see in the 
top panel that the average amplitude increases as a function of $Y_s$, as expected. However, there is a large vertical spread due to the spread in 
the mass, as revealed by the colour coding. This is also expected, as $\langle A_\nu \rangle$ is known to depend on $Y_s$ as well as on $M$ 
\citep{verm14b}. To filter out the models with extreme masses, we defined a cost function,
\begin{equation}
\chi^2 = \chi^2_{\rm Teff} + \chi^2_{[{\rm Fe}/{\rm H}]_s} + \chi^2_{02} + \chi^2_{01} + \chi^2_{10},
\label{chi2}
\end{equation}
where the last three terms are the reduced chi-squares for the different frequency ratios computed using their covariance matrices. The covariances 
were calculated using 10000 realizations of the observed oscillation frequencies. We would ideally expect the $\chi^2$ to be close to 5 for the 
best-fitting models of the stars. 

We can see from the middle panel of Figure~\ref{fig1} that the models with the extreme masses on the top and bottom have very large values of 
$\chi^2$ (note the logarithmic scale). Such models do not represent 16 Cyg A, and should either be dropped or given less weight while fitting the 
straight line for the calibration. We do both: the models with $\chi^2$ greater than a threshold are dropped first, and then a straight-line is 
fitted to the remaining models with the weight of $e^{-\chi^2/5}$. We wish to point out that the threshold on $\chi^2$ is not important for $Y_s$ 
determination because of the exponential weighting, and is applied only to see the correlation clearly. Note that this process of filtering is 
slightly different from that used in \citet{verm14a}, where we first determined the age using the spectroscopic and seismic data, and then filtered 
the models that had age within $1\sigma$ of the determined value. The filtering method used in the current analysis is more generic -- using only 
the observables -- and is less model dependent. As we shall see in Section~\ref{helium_abundances}, the surface helium abundances found in the 
current work for 16 Cyg A \& B are in good agreement with the values obtained by \citet{verm14a}. 

\begin{figure*}
\includegraphics[width=\textwidth]{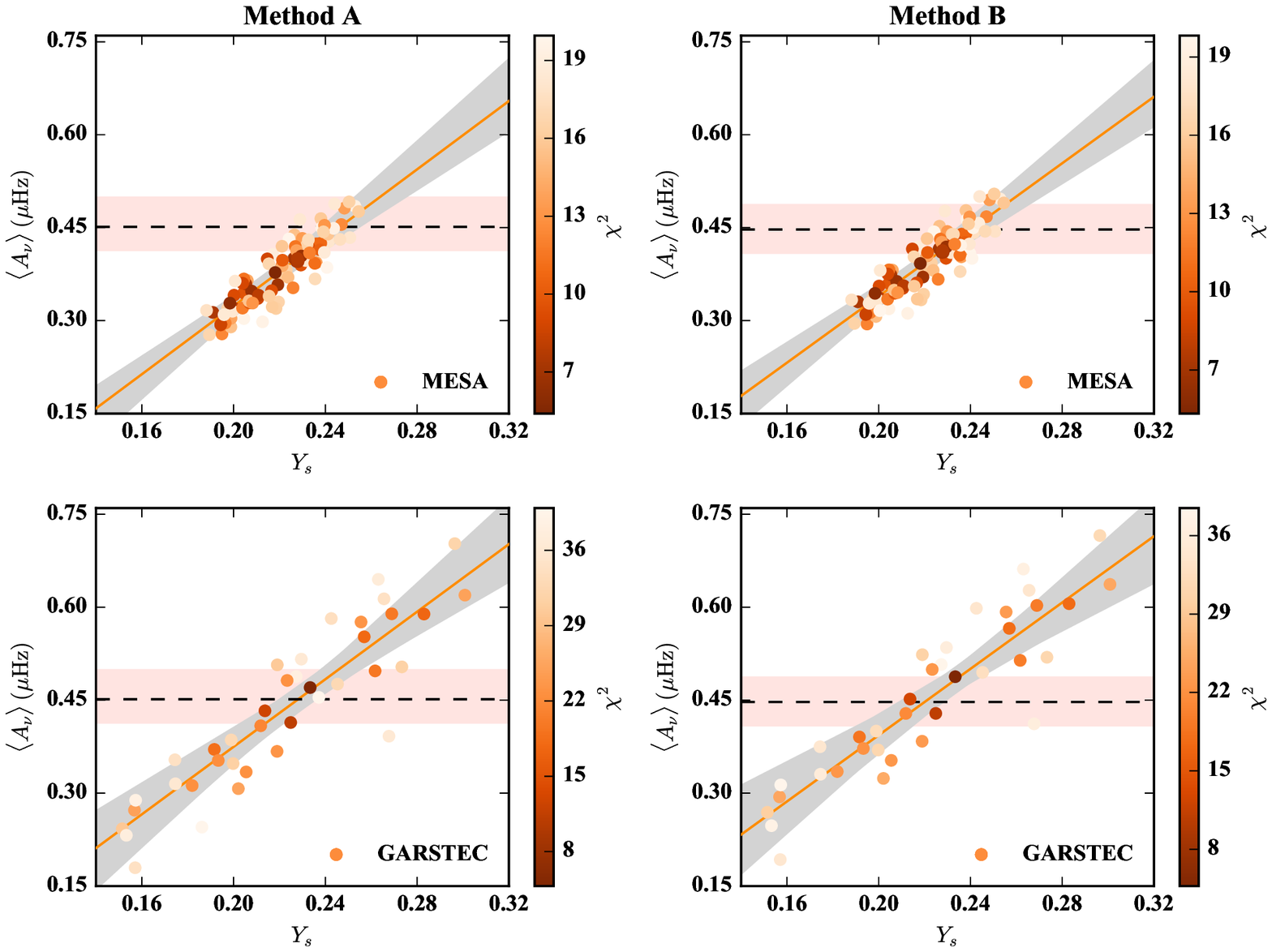}
\caption{Calibration of the observed average amplitude against the surface helium abundance for 16 Cyg A. The two columns represent the results 
obtained using the two different fitting methods, while the two rows show the two different sets of the calibration models. The dashed line 
represents the observed average amplitude with $1\sigma$ uncertainty shown by the shaded region around it. The points in a panel represent the 
average amplitude for the models with the colour indicating their $\chi^2$ value (as defined in Eq.~\ref{chi2}). The continuous line is a weighted 
straight-line fit to the points with the band around it representing the Monte-Carlo regression uncertainty.}
\label{fig2}
\end{figure*}

\begin{figure*}
\includegraphics[width=\textwidth]{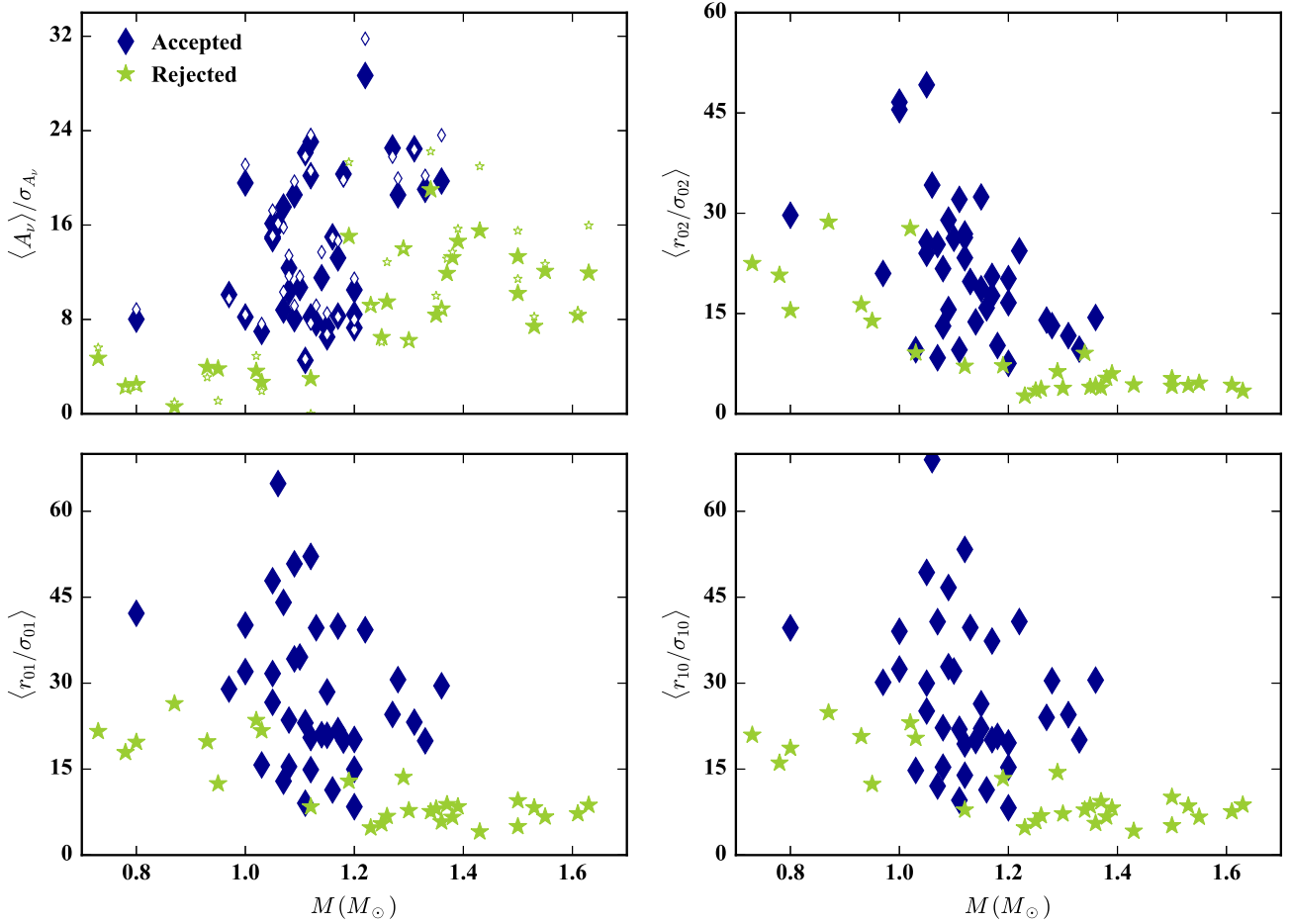}
\caption{Signal-to-noise ratio for the various observables as a function of the mass \citep[mass is taken from][]{verm17}. The top left panel shows 
the S/N of the average amplitude for all the stars in the LEGACY sample. The big-filled and small-open symbols in that panel are obtained using 
the fitting Methods A and B, respectively. The rest of the panels show the S/N of the three ratios averaged over the radial order. The diamond 
points represent the 38 stars for which both the requirements are met (see the text), while the star symbols represent those for which at least one 
requirement is not fulfilled.}
\label{fig3}
\end{figure*}

Figure~\ref{fig2} shows four calibration diagrams for 16 Cyg A obtained using the Methods A and B and the models MESA and GARSTEC. We can see that 
the two columns look very similar meaning that the extraction of the glitch signature is insensitive to the method used. Figure~\ref{fig2} can be 
used to get four estimates of $Y_s$ for 16 Cyg A. Note in the figure that, to determine the surface helium abundance of a star reliably, we need 
(1) a precise enough determination of the observed $\langle A_\nu \rangle$, and (2) a reasonably tight correlation. We have 38 stars in the LEGACY 
sample for which both the requirements are met. Unfortunately, there is no straightforward way to relate the above two requirements with the 
quality of the observations. Having said that, Figure~\ref{fig3} makes an attempt to explain why the rest of the 28 stars from the LEGACY sample 
are rejected. 

The precision of the observed $\langle A_\nu \rangle$ not only depends on the precision of the observed oscillation frequencies, but also on the 
number of modes detected and on the mass of the star. The mass dependence is due to the fact that the strength of the helium glitch signature 
depends on the mass -- the larger the mass, the stronger the signature -- making it easier to detect \citep{verm14b}. Typically, the observed power 
spectrum of a low-mass star has poor S/N, and hence the number of detected modes are limited \citep[see][]{appo12,lund17}, resulting in a low S/N 
$\langle A_\nu \rangle$ as seen in Figure~\ref{fig3}. Note however that the measured oscillation frequencies of such stars are very precise 
because of the narrow linewidths, and the resulting frequency ratios have large S/N. On the other hand, the linewidths for the high-mass stars are 
very broad and the oscillation frequencies have large errorbars. Although this is not a problem for the $\langle A_\nu \rangle$ because of the 
strong helium signature arising from such stars, the frequency ratios have poor S/N (see Figure~\ref{fig3}).  

\begin{table*}
\centering
\scriptsize
\caption{Envelope helium abundance of all the 38 stars obtained using the two different fitting methods (A and B) and two sets of calibration models 
(MESA and GARSTEC). The last two columns are the settling of the helium and heavy elements obtained using the best-fitting MESA models.}
\label{tab3}
\begin{tabular}{cccccccc}
\hline\hline
  KIC  &  \multicolumn{2}{c}{Method A}  &  &  \multicolumn{2}{c}{Method B} & $\delta Y = Y_i - Y_s$ & $\delta Z = Z_i - Z_s$\\
\cline{2-3} \cline{5-6}\\
       &  MESA  &  GARSTEC  &  &  MESA  &  GARSTEC\\
\hline
   3427720  &  $0.205_{-0.014}^{+0.015}$  &  $0.193_{-0.015}^{+0.016}$  &  &  $0.201_{-0.012}^{+0.014}$  &  $0.191_{-0.013}^{+0.015}$  &  0.0262  &  0.0016\\
   3632418  &  $0.304_{-0.018}^{+0.016}$  &  $0.248_{-0.010}^{+0.009}$  &  &  $0.313_{-0.019}^{+0.017}$  &  $0.249_{-0.010}^{+0.010}$  &  0.0717  &  0.0041\\
   3656476  &  $0.271_{-0.018}^{+0.024}$  &  $0.267_{-0.022}^{+0.029}$  &  &  $0.250_{-0.014}^{+0.019}$  &  $0.243_{-0.017}^{+0.024}$  &  0.0368  &  0.0043\\
   3735871  &  $0.217_{-0.014}^{+0.016}$  &  $0.206_{-0.016}^{+0.019}$  &  &  $0.215_{-0.014}^{+0.017}$  &  $0.204_{-0.017}^{+0.020}$  &  0.0136  &  0.0009\\
   4914923  &  $0.242_{-0.013}^{+0.012}$  &  $0.222_{-0.009}^{+0.008}$  &  &  $0.240_{-0.015}^{+0.014}$  &  $0.220_{-0.010}^{+0.009}$  &  0.0411  &  0.0033\\
   5184732  &  $0.291_{-0.026}^{+0.028}$  &  $0.313_{-0.031}^{+0.033}$  &  &  $0.293_{-0.028}^{+0.027}$  &  $0.320_{-0.034}^{+0.033}$  &  0.0345  &  0.0043\\
   6106415  &  $0.222_{-0.012}^{+0.014}$  &  $0.203_{-0.009}^{+0.011}$  &  &  $0.218_{-0.013}^{+0.014}$  &  $0.201_{-0.010}^{+0.011}$  &  0.0401  &  0.0028\\
   6116048  &  $0.227_{-0.011}^{+0.011}$  &  $0.206_{-0.008}^{+0.008}$  &  &  $0.228_{-0.011}^{+0.010}$  &  $0.205_{-0.008}^{+0.008}$  &  0.0560  &  0.0029\\
   6225718  &  $0.235_{-0.009}^{+0.009}$  &  $0.227_{-0.009}^{+0.009}$  &  &  $0.235_{-0.010}^{+0.009}$  &  $0.229_{-0.010}^{+0.009}$  &  0.0413  &  0.0027\\
   6603624  &  $0.241_{-0.017}^{+0.018}$  &  $0.221_{-0.007}^{+0.008}$  &  &  $0.238_{-0.018}^{+0.021}$  &  $0.221_{-0.007}^{+0.008}$  &  0.0388  &  0.0046\\
   6933899  &  $0.223_{-0.010}^{+0.009}$  &  $0.194_{-0.008}^{+0.008}$  &  &  $0.218_{-0.009}^{+0.009}$  &  $0.193_{-0.008}^{+0.008}$  &  0.0346  &  0.0025\\
   7296438  &  $0.247_{-0.023}^{+0.025}$  &  $0.233_{-0.018}^{+0.019}$  &  &  $0.262_{-0.021}^{+0.030}$  &  $0.245_{-0.017}^{+0.024}$  &  0.0434  &  0.0039\\
   7510397  &  $0.261_{-0.016}^{+0.018}$  &  $0.209_{-0.007}^{+0.008}$  &  &  $0.251_{-0.016}^{+0.017}$  &  $0.210_{-0.008}^{+0.009}$  &  0.0727  &  0.0042\\
   7680114  &  $0.198_{-0.011}^{+0.018}$  &  $0.186_{-0.011}^{+0.018}$  &  &  $0.197_{-0.009}^{+0.014}$  &  $0.185_{-0.010}^{+0.016}$  &  0.0437  &  0.0040\\
   7940546  &  $0.331_{-0.028}^{+0.025}$  &  $0.222_{-0.013}^{+0.012}$  &  &  $0.328_{-0.025}^{+0.023}$  &  $0.229_{-0.014}^{+0.013}$  &  0.0809  &  0.0048\\
   8006161  &  $0.246_{-0.020}^{+0.025}$  &  $0.258_{-0.023}^{+0.029}$  &  &  $0.224_{-0.019}^{+0.025}$  &  $0.231_{-0.022}^{+0.028}$  &  0.0208  &  0.0023\\
   8179536  &  $0.270_{-0.033}^{+0.038}$  &  $0.280_{-0.035}^{+0.040}$  &  &  $0.264_{-0.034}^{+0.040}$  &  $0.270_{-0.035}^{+0.041}$  &  0.0594  &  0.0043\\
   8228742  &  $0.230_{-0.019}^{+0.020}$  &  $0.207_{-0.010}^{+0.011}$  &  &  $0.237_{-0.018}^{+0.018}$  &  $0.211_{-0.010}^{+0.010}$  &  0.0617  &  0.0035\\
   8379927  &  $0.250_{-0.008}^{+0.008}$  &  $0.237_{-0.008}^{+0.008}$  &  &  $0.251_{-0.008}^{+0.009}$  &  $0.237_{-0.008}^{+0.008}$  &  0.0254  &  0.0014\\
   8394589  &  $0.253_{-0.022}^{+0.025}$  &  $0.215_{-0.016}^{+0.018}$  &  &  $0.257_{-0.023}^{+0.021}$  &  $0.219_{-0.017}^{+0.016}$  &  0.0561  &  0.0024\\
   8694723  &  $0.251_{-0.017}^{+0.016}$  &  $0.194_{-0.015}^{+0.014}$  &  &  $0.250_{-0.015}^{+0.016}$  &  $0.197_{-0.013}^{+0.014}$  &  0.0860  &  0.0028\\
   8760414  &  $0.203_{-0.019}^{+0.022}$  &  $0.178_{-0.009}^{+0.011}$  &  &  $0.201_{-0.018}^{+0.018}$  &  $0.176_{-0.009}^{+0.009}$  &  0.0695  &  0.0012\\
   8938364  &  $0.234_{-0.010}^{+0.010}$  &  $0.220_{-0.009}^{+0.010}$  &  &  $0.227_{-0.010}^{+0.010}$  &  $0.216_{-0.009}^{+0.010}$  &  0.0603  &  0.0031\\
   9098294  &  $0.256_{-0.016}^{+0.015}$  &  $0.247_{-0.016}^{+0.016}$  &  &  $0.243_{-0.014}^{+0.017}$  &  $0.231_{-0.014}^{+0.018}$  &  0.0457  &  0.0026\\
   9139151  &  $0.233_{-0.014}^{+0.014}$  &  $0.216_{-0.015}^{+0.015}$  &  &  $0.230_{-0.012}^{+0.014}$  &  $0.213_{-0.013}^{+0.014}$  &  0.0196  &  0.0013\\
   9410862  &  $0.221_{-0.021}^{+0.021}$  &  $0.206_{-0.018}^{+0.018}$  &  &  $0.211_{-0.021}^{+0.016}$  &  $0.198_{-0.019}^{+0.014}$  &  0.0528  &  0.0022\\
   9965715  &  $0.312_{-0.053}^{+0.061}$  &  $0.253_{-0.027}^{+0.032}$  &  &  $0.273_{-0.046}^{+0.050}$  &  $0.235_{-0.023}^{+0.026}$  &  0.0871  &  0.0041\\
  10068307  &  $0.257_{-0.015}^{+0.013}$  &  $0.251_{-0.011}^{+0.010}$  &  &  $0.262_{-0.011}^{+0.013}$  &  $0.249_{-0.009}^{+0.010}$  &  0.0675  &  0.0029\\
  10079226  &  $0.257_{-0.031}^{+0.032}$  &  $0.244_{-0.030}^{+0.032}$  &  &  $0.252_{-0.027}^{+0.034}$  &  $0.240_{-0.027}^{+0.035}$  &  0.0262  &  0.0017\\
  10162436  &  $0.324_{-0.022}^{+0.021}$  &  $0.273_{-0.016}^{+0.015}$  &  &  $0.324_{-0.022}^{+0.022}$  &  $0.273_{-0.016}^{+0.016}$  &  0.0708  &  0.0039\\
  10454113  &  $0.279_{-0.015}^{+0.017}$  &  $0.276_{-0.016}^{+0.018}$  &  &  $0.278_{-0.015}^{+0.018}$  &  $0.273_{-0.016}^{+0.019}$  &  0.0483  &  0.0029\\
  10516096  &  $0.212_{-0.012}^{+0.012}$  &  $0.204_{-0.011}^{+0.011}$  &  &  $0.204_{-0.010}^{+0.010}$  &  $0.198_{-0.010}^{+0.010}$  &  0.0555  &  0.0033\\
  10644253  &  $0.270_{-0.020}^{+0.026}$  &  $0.256_{-0.020}^{+0.026}$  &  &  $0.270_{-0.022}^{+0.030}$  &  $0.257_{-0.021}^{+0.029}$  &  0.0145  &  0.0010\\
  10963065  &  $0.233_{-0.020}^{+0.022}$  &  $0.210_{-0.016}^{+0.019}$  &  &  $0.221_{-0.016}^{+0.020}$  &  $0.199_{-0.013}^{+0.017}$  &  0.0525  &  0.0030\\
  12009504  &  $0.256_{-0.029}^{+0.029}$  &  $0.227_{-0.018}^{+0.019}$  &  &  $0.239_{-0.024}^{+0.023}$  &  $0.222_{-0.016}^{+0.016}$  &  0.0713  &  0.0045\\
  12069424  &  $0.246_{-0.014}^{+0.017}$  &  $0.228_{-0.014}^{+0.017}$  &  &  $0.240_{-0.014}^{+0.015}$  &  $0.220_{-0.014}^{+0.015}$  &  0.0404  &  0.0033\\
  12069449  &  $0.255_{-0.010}^{+0.010}$  &  $0.225_{-0.008}^{+0.008}$  &  &  $0.248_{-0.009}^{+0.009}$  &  $0.219_{-0.007}^{+0.007}$  &  0.0371  &  0.0032\\
  12258514  &  $0.250_{-0.008}^{+0.007}$  &  $0.215_{-0.008}^{+0.007}$  &  &  $0.250_{-0.007}^{+0.007}$  &  $0.213_{-0.007}^{+0.007}$  &  0.0471  &  0.0034\\
\hline
\end{tabular}
\end{table*}

\begin{figure}
\includegraphics[width=0.9\columnwidth]{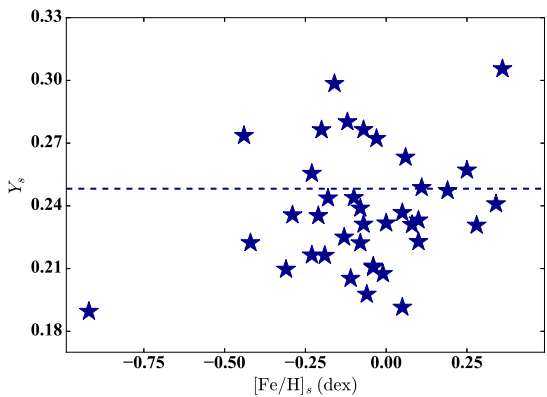}
\caption{Surface helium abundance as a function of the surface metallicity for all the 38 stars. The points represent the centre of the helium range 
spanned by the four estimates listed in Table~\ref{tab3}. The horizontal dashed line represents the primordial helium abundance.}
\label{fig4}
\end{figure}

Assuming that the tightness of the correlation in the calibration diagram depends on the S/N of the frequency ratios, the rejection of all the 28 
stars can be understood from Figure~\ref{fig3}. The low-mass stars are rejected primarily because of the poor S/N of $\langle A_\nu \rangle$, while 
the high-mass stars are rejected because of the poor S/N of the frequency ratios. It should be noted that the overshoot plays an important 
role together with the mass in determining the size of the convective core for the high-mass stars. Since we select the calibration models based 
on the frequency ratios, which puts constraint on the convective core size, there is a degeneracy between the mass and overshoot -- the larger the
overshoot, the smaller the mass. This trade-off between the mass and overshoot results calibration models with significantly different masses, 
adding more scatter in the calibration diagram for the high-mass stars.

\subsection{Surface helium abundance}
\label{helium_abundances}
The surface helium abundance for all the 38 stars obtained using the different fitting methods and different sets of calibration models are listed 
in Table~\ref{tab3}. The centre of the range spanned by the four estimates of $Y_s$ is shown in Figure~\ref{fig4} as a function of the surface 
metallicity. Note in the figure that a large fraction of the stars have $Y_s$ below the standard Big Bang nucleosynthesis value 
\citep[$0.2482\pm0.0007$;][]{stei10}, confirming the significant settling of the helium in these stars as is known to happen in the Sun. The table 
also lists the settling of the helium, $\delta Y = Y_i - Y_s$, and heavy elements, $\delta Z = Z_i - Z_s$, predicted by the best-fitting MESA models.
This will be used in Section~\ref{enrichment} to get the preliminary estimates of the initial abundances. 
 
\begin{figure*}
\includegraphics[width=\textwidth]{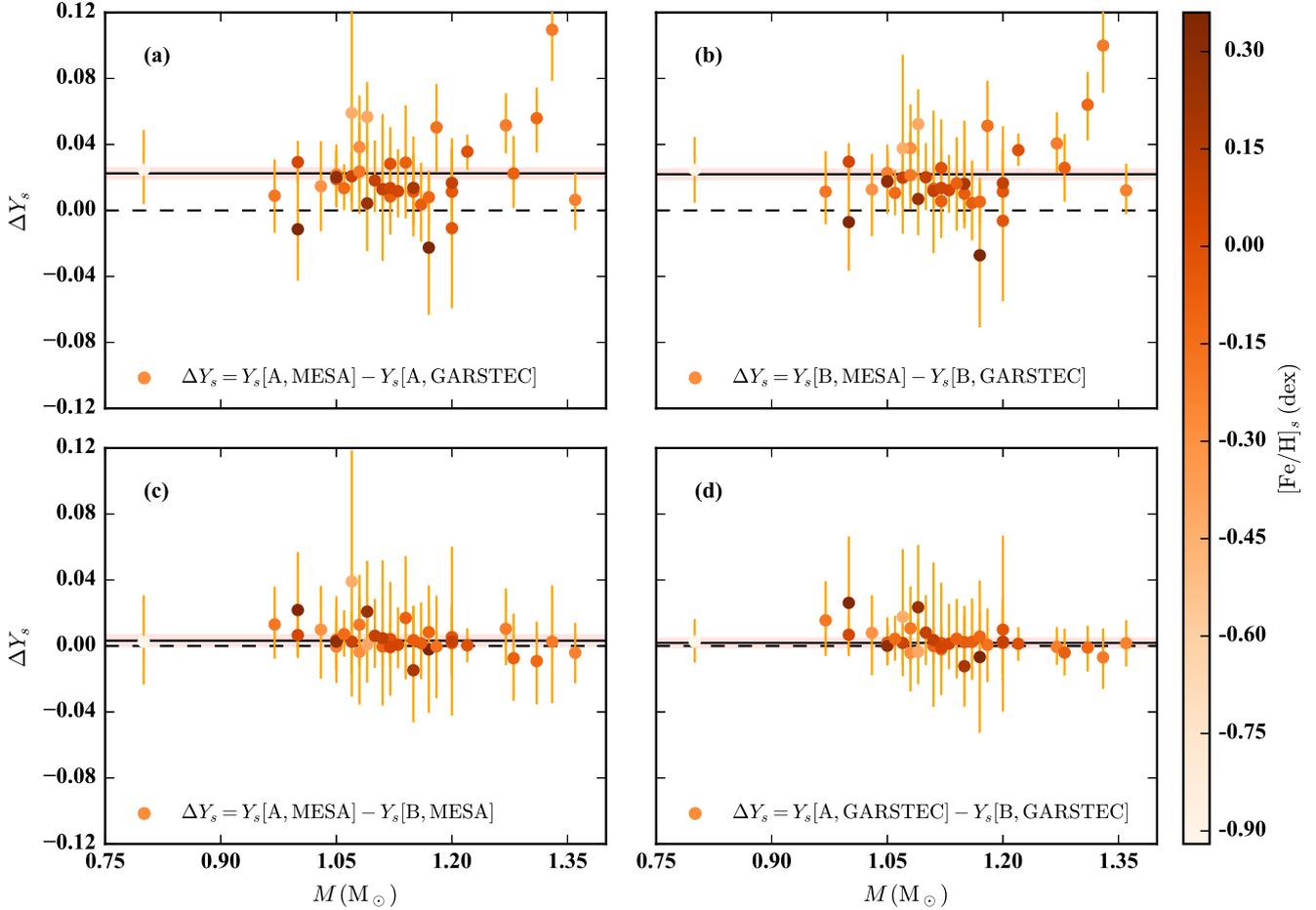}
\caption{Comparison of the different estimates of the envelope helium abundance for all the 38 stars. The panels (a--b) compare the estimates
obtained using the different sets of calibration models (MESA and GARSTEC) for a given fitting method, while the panels (c--d) compare the estimates 
obtained using the different fitting methods (A and B) for a given set of calibration models. The horizontal dashed line indicates the zero level. 
The continuous line is the weighted average of the points with the band around it representing the $1\sigma$ uncertainty. The colour represents 
the metallicity of the stars.} 
\label{fig5}
\end{figure*}

Figure~\ref{fig5} shows comparisons among the different estimates of $Y_s$. As we can see from the bottom panels, the values obtained using the 
two different fitting methods agree well within 1$\sigma$. For this reason and for the sake of clarity, we shall present the results only from the 
Method A in the subsequent sections. The envelope helium abundances obtained using the different sets of calibration models also agree within 
1$\sigma$ for most of the stars as seen in the top panels of Figure~\ref{fig5}, except for a few high-mass stars. There is, however, a noticeable 
systematic shift of approximately 0.02, with the MESA models giving systematically larger values of $Y_s$ than the GARSTEC models. This offset 
is due to the differences in the input physics used in the two sets of models. We shall show in Section~\ref{yrec_results} using the YREC models 
that this small systematic difference is a result of the fact that the MESA includes atomic diffusion and the GARSTEC does not.

We have 3 solar-type stars with previously determined $Y_s$: the Sun, 16 Cyg A and 16 Cyg B. The solar helium abundance was determined using the 
intermediate degree oscillation frequencies \citep[$0.248\pm0.003$;][]{basu98}. The $Y_s$ for the binary system 16 Cyg A \& B was estimated using 
the helium glitch signature in the low degree oscillation frequencies in the way it is determined in the current work (minor differences 
lie in the details of the methods and in the length of the time series used to compute the observed oscillation frequencies). \citet{verm14a} 
found the $Y_s$ for 16 Cyg A to be in the range 0.231--0.251 and for 16 Cyg B in the range 0.218--0.266. We can see from Table~\ref{tab3} that the 
different estimates of $Y_s$ for 16 Cyg A (KIC 12069424) \& B (KIC 12069449) are in good agreement, including the previous 
determinations. 

We used Sun-as-a-star data from \citet{lund17} to determine the solar helium abundance using the Method A. We found $Y_s$ to be 
$0.240_{-0.015}^{+0.025}$ and $0.223_{-0.013}^{+0.022}$ for the MESA and GARSTEC calibration models, respectively. Note that the 
estimate obtained using the MESA is again larger than the one obtained using the GARSTEC by roughly the same amount ($0.02$). Since the $Y_s$ 
obtained using the MESA models is closer to the helioseismic value, we may expect the values derived using the MESA models for the other stars with 
masses and metallicities similar to the Sun to be more accurate than the values obtained using the GARSTEC models. This is also supported by 
the fact that the diffusion models are more realistic than the non-diffusion ones for such stars. 

Element settling in relatively massive stars (mass can be lower than 1.2 M$_\odot$ for a metallicity of -0.5 dex) is not very well understood. 
Models of atomic diffusion predict excessive settling of the helium and heavy elements in the envelope of such stars \citep{more02}. This does not 
mean that the models of atomic diffusion are wrong, rather it points towards the importance of the non-standard processes that compete with atomic 
diffusion \citep[see e.g.][also see Section~\ref{enrichment} for further discussion and references]{turc98}. For such stars, the non-diffusion 
models are expected to be closer to the real stars in terms of reproducing the observed surface abundances (unless the non-standard processes are 
better understood, and are included in the stellar model calculations along with atomic diffusion), hence the $Y_s$ obtained using the non-diffusion 
GARSTEC models are expected to be more accurate. The range of $Y_s$ obtained using the MESA and GARSTEC models provides the largest possible 
systematic uncertainty expected from the uncertainties in the chemical element transport in the envelope, as these two sets of models represent the 
two extreme cases of the helium settling.

\begin{table*}
\centering
\scriptsize
\caption{Envelope helium abundance of four stars obtained using the Method A and both sets of YREC models. The last two columns are the settling 
of the helium and heavy elements obtained using the best-fitting YREC (diffusion) models.}
\label{tab4}
\begin{tabular}{ccccc}
\hline\hline
  Star  &  YREC (dif.)  &  YREC (nodif.)  &  $\delta Y = Y_i - Y_s$  &  $\delta Z = Z_i - Z_s$\\
\hline
       16 Cyg A  &  0.245$_{-0.019}^{+0.024}$  &  0.232$_{-0.014}^{+0.017}$  &  0.0373  &  0.0038\\
       16 Cyg B  &  0.274$_{-0.016}^{+0.016}$  &  0.245$_{-0.009}^{+0.009}$  &  0.0389  &  0.0034\\
    KIC 6106415  &  0.223$_{-0.012}^{+0.013}$  &  0.211$_{-0.012}^{+0.014}$  &  0.0389  &  0.0029\\
    KIC 6116048  &  0.223$_{-0.011}^{+0.011}$  &  0.210$_{-0.012}^{+0.012}$  &  0.0545  &  0.0027\\
\hline
\end{tabular}
\end{table*}

\begin{figure*}
\includegraphics[width=\textwidth]{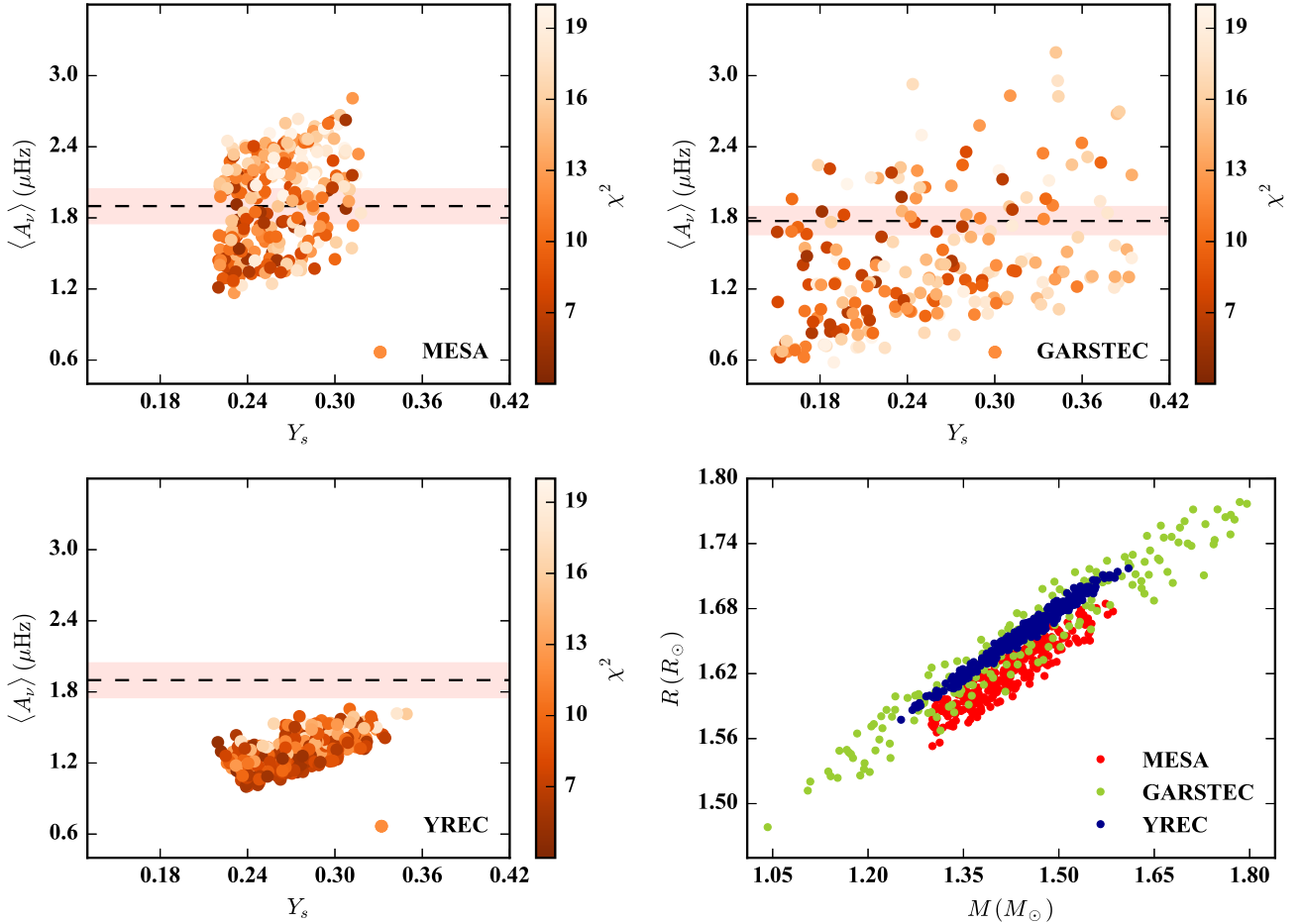}
\caption{Calibration diagrams (the top panels and the bottom left panel) for KIC 2837475 obtained using the Method A (see the caption for 
Figure~\ref{fig2}). The corresponding models are also shown in a mass-radius diagram in the bottom right panel.}
\label{fig6}
\end{figure*}

\subsection{Impact of the atomic diffusion}
\label{yrec_results}
To demonstrate the source of the systematic shift of about 0.02 seen in the top panels of Figure~\ref{fig5}, we constructed two sets of YREC models
differing only in the diffusion (one includes atomic diffusion and other does not). We used the fitting Method A for both the sets of models to 
determine $Y_s$ for 16 Cyg A \& B, KIC 6106415 and 6116048. We can see in Table~\ref{tab4} that these estimates of $Y_s$ agree well with those 
presented in Table~\ref{tab3}, pointing towards the robustness of $Y_s$ determination. We can also see that the envelope helium abundance obtained 
using the diffusion models are systematically larger than those found with the non-diffusion models by about 0.02, confirming the fact that the 
systematic shift is indeed a result of the uncertainties in the chemical element transport in the stellar interior. Table~\ref{tab4} also lists the 
settling predicted by the best-fitting YREC (diffusion) models, and can be compared with those obtained with the MESA (see Table~\ref{tab3}). A good 
level of agreement reassures the robustness of $\delta Y$ and $\delta Z$ determinations against the uncertainties in the stellar properties and the
input physics. This however does not necessarily mean that the estimates are accurate, because the largest uncertainty on $\delta Y$ and $\delta Z$ 
results from the neglect of the various physical processes that compete with the atomic diffusion, which have been ignored in both the sets of 
models.

We wish to take here an example of YREC models to point out one caveat related to the determination of $Y_s$ through the model calibration. Every 
choice of the vector ($M$, $Y_i$, $[{\rm Fe}/{\rm H}]_i$, $\alpha_{\rm MLT}$, $\alpha_{\rm OV}$) for different values of $M$, $Y_i$, 
$[{\rm Fe}/{\rm H}]_i$, $\alpha_{\rm MLT}$ and $\alpha_{\rm OV}$ results a point in the calibration diagram. This means that, in principle, it is 
possible to choose a set of these vectors such that the calibration diagram has tight artificial correlation irrespective of the observed data 
quality (essentially disregarding systematically some of the vectors that lead to off-the-trend points in the calibration diagram). This may also 
happen in practice, which is illustrated below for KIC 2837475. KIC 2837475 is one of the 28 rejected stars with the poor S/N frequency ratios 
($\langle r_{02} / \sigma_{02} \rangle = 4.1$, $\langle r_{01} / \sigma_{01} \rangle = 5.0$ and $\langle r_{10} / \sigma_{10} \rangle = 5.2$).

Figure~\ref{fig6} shows the calibration diagrams for KIC 2837475. Note that this star has relatively high-mass ($\approx 1.5$ M$_\odot$), 
and hence the corresponding models in the figure exclude the atomic diffusion in all the three sets. We can see in the figure that, despite the 
poor S/N of the frequency ratios, the YREC models show reasonably tight correlation in the calibration diagram. On the other hand, the MESA and 
GARSTEC models show much larger scatter. This is partly due to the fact that the YREC models by construction have $T_{\rm eff}$ and 
$[{\rm Fe}/{\rm H}]_s$ within $1\sigma$ of the observations. More importantly, however, this is a result of using the semi-empirical scaling 
relations \citep{kjel95} to constrain the initial parameter space. We can see in the mass-radius diagram that the YREC models are just a subset of 
the MESA and GARSTEC models, showing much tighter relation between the mass and radius. 

For the stars with the mass and metallicity similar to the Sun, like the ones listed in Table~\ref{tab4}, the scaling relations are known to work 
well, and hence resulting in consistent values for the $Y_s$. However, for the relatively high-mass stars, like the one shown in Figure~\ref{fig6}, 
the artificial correlation can provide highly biased estimate of $Y_s$. To reduce the possibility of introducing misleading tight correlations in 
the calibration diagrams, we considered all the relevant parameters free including the mixing-length and the overshoot for both the MESA and GARSTEC. 
Furthermore, we sampled the initial parameter space uniformly, and selected the calibration models solely based on the observations.

\subsection{Galactic enrichment ratio}
\label{enrichment}
The determination of the enrichment ratio requires the initial values of the helium and metal mass fractions, which rely on the measurements of the
corresponding surface values as well as on the models of the chemical element transport in the stellar envelope. The inclusion of the atomic 
diffusion in the solar models reduced significantly the discrepancies between the model and helioseismic data \citep[see e.g.,][]{jcd93b,guen96}, 
pointing towards the importance of this process in the solar interior. However, the same models of the atomic diffusion predict complete depletion of
the helium and heavy elements in the envelope of the stars of masses approximately greater than 1.4 M$_\odot$ (also depends on the metallicity and 
age). This is unrealistic given the measurements of the heavy element abundances in the A and F-type stars \citep[see e.g.][]{vare99} and the 
detections of the helium glitch signature in the F-type stars \citep[see][]{verm17}. Additional processes like radiative levitation, mass 
loss, turbulence etc. can potentially reduce the settling \citep[see e.g.][]{vauc78a,vauc78b,vauc99,rich00}, and lead to a better agreement between 
the observed and model predicted surface abundances \citep[see e.g.][]{rich00,thea01,mich11,cast16}. A detailed investigation of the settling 
including the microscopic diffusion, radiative levitation and other processes is beyond the scope of this work. 

\begin{figure*}
\includegraphics[width=\textwidth]{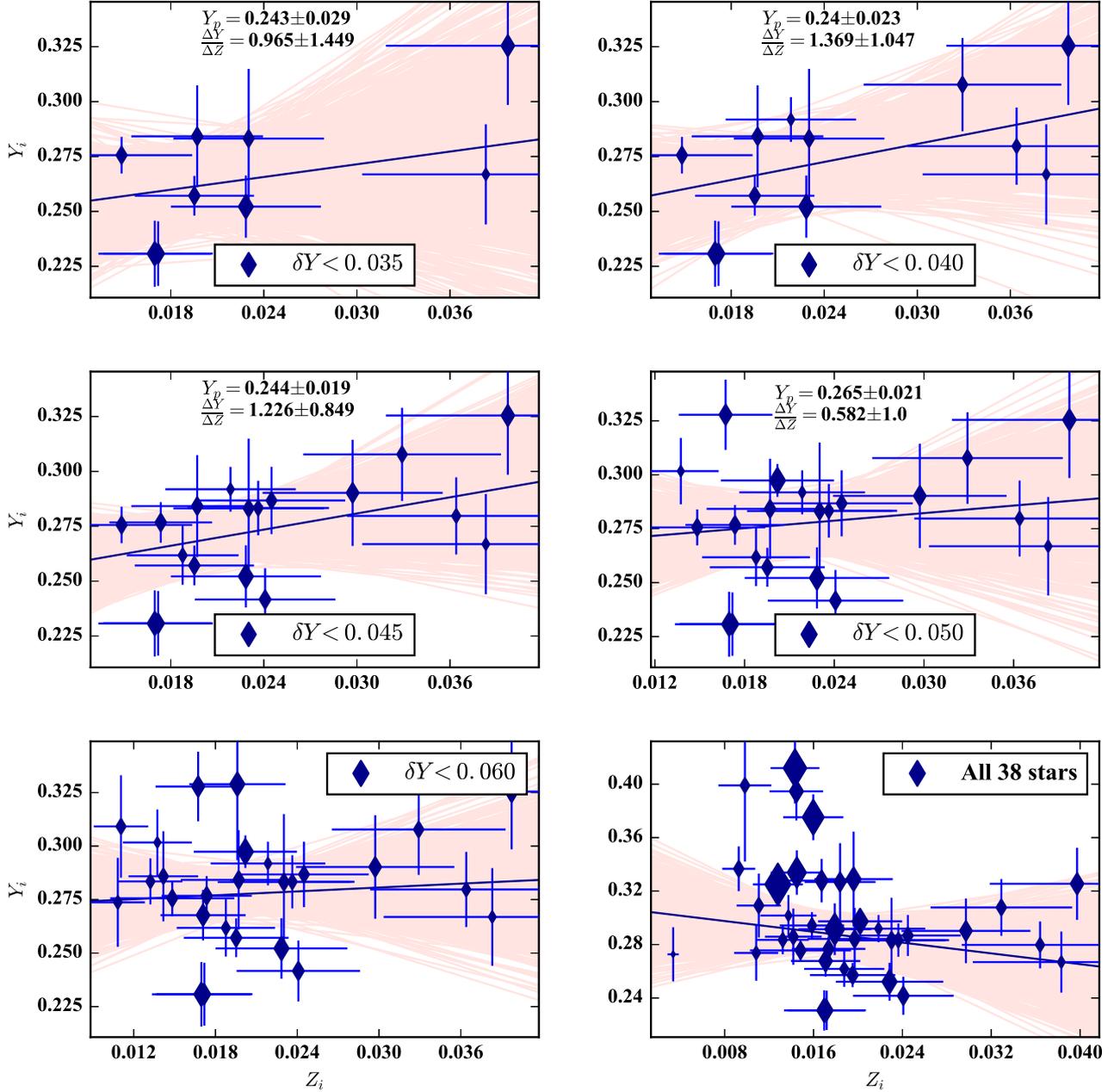}
\caption{Initial helium mass fraction as a function of the initial metal mass fraction. Different panels show the stars with the helium settling
below a threshold. The points are exponentially scaled with the mass. The blue line is a weighted straight-line fit to the points with the band 
around it representing the Monte-Carlo regression uncertainty. The intercept ($Y_p$) and the slope ($\Delta Y / \Delta Z$) of the fitted line is 
shown in the top four panels.}
\label{fig7}
\end{figure*}

The point of the above discussion is that the excessive settling of the helium and heavy elements predicted by the standard stellar models with 
atomic diffusion (excluding the non-standard processes) for the relatively massive and metal poor stars is unrealistic. Hence the estimated 
$\delta Y$ and $\delta Z$ as well as $Y_i$ and $Z_i$ are inaccurate for such stars. For this reason, to determine the enrichment ratio we shall 
consider only those stars for which the stellar models predict helium settling below a threshold (the mass alone is not a good quantity for this 
purpose because the settling also depends substantially on the metallicity and evolutionary state). 

We estimated the initial abundances from the measured surface values using the settling from the best-fitting MESA models. Figure~\ref{fig7} 
shows $Y_i$ against $Z_i$ with the different thresholds on the helium settling. We fitted a straight line to $Y_i$ as a function of $Z_i$ using the 
weighted least squares method to get the ``preliminary" estimates of the primordial helium abundance (the intercept, $Y_p$) and the galactic 
enrichment ratio (the slope, $\Delta Y / \Delta Z$). Note that only the uncertainty on $Y_i$ is currently used as weight in producing the 
linear relationship. The reason is that there are only a few data points at high $Z_i$ with rather large errorbars, and hence including 
the uncertainty on $Z_i$ in the fit makes the slope and intercept highly uncertain. Moreover, the inclusion of this uncertainty makes the 
optimization process nonlinear, and fits suffer from convergence issues due to poor constraint on the slope and intercept. We made two unsuccessful 
attempts to fit the data with uncertainties in both $Y_i$ and $Z_i$ using {\it bivariate correlated errors and intrinsic scatter} 
\citep[BCES;][]{akri96,nemm12}\footnote{https://github.com/rsnemmen/BCES} and {\it affine invariant markov chain monte carlo ensemble sampler} 
\citep[emcee: The MCMC Hammer;][]{good10,fore13}\footnote{http://dfm.io/emcee/current/}. Clearly we need either more data points at high $Z_i$ or 
more precise measurements of the surface metallicity to perform a proper fitting including the uncertainties in both $Y_i$ and $Z_i$. In the 
foreseeable future, we do not expect much improvement in the precision of the metallicity measurements, however the sample size is anticipated to get 
much bigger in the {\it Planetary Transits and Oscillations of stars} \citep[PLATO;][]{raue14} era. To simulate larger sample size, we duplicated 
the data points with $Z_i > 0.025$ (the data points with the largest $Z_i$ errorbars), and tried fitting the straight line with errors in both $Y_i$ 
and $Z_i$ using emcee. Although results are meaningless due to artificially added data points, we confirm that the emcee chains do converge now. 
Therefore larger sample size in the future can potentially alleviate the problem of poor precision of $Z_i$.

We note from the legends in Figure~\ref{fig7} the precision of $Y_p$ and $\Delta Y / \Delta Z$ as a function of the threshold on $\delta Y$. The 
precision initially increases until $\delta Y = 0.045$ because of the increase in the number of the data points. However, as we further increase the 
threshold, the precision drops and the trend becomes meaningless eventually when including all the 38 stars. We attribute this behaviour to the 
unreliable estimates of $Y_i$ and $Z_i$ for the stars with $\delta Y > 0.045$. For this reason, we recommend the estimates of $Y_p = 0.244 \pm 0.019$
and $\Delta Y / \Delta Z = 1.226 \pm 0.849$ from the middle left panel with $\delta Y < 0.045$. To test the robustness of the results, we also 
fitted a straight line to $Z_i$ as a function of $Y_i$ using weighted least squares method (weight being the uncertainty on $Z_i$), and inverted the 
relationship to get $Y_p$ and $\Delta Y / \Delta Z$. For the panel corresponding to $\delta Y < 0.045$, we find $Y_p = 0.094 \pm 0.128$ and 
$\Delta Y / \Delta Z = 8.019 \pm 3.618$. Clearly the uncertainties on the estimates are very large due to the large uncertainty on $Z_i$. We
wish to further point out that the quoted uncertainty on $Z_i$ represents only the statistical uncertainty. It may also have systematic uncertainties 
arising from the uncertainties in the models of stellar atmosphere (which affects the measurement of $[{\rm Fe}/{\rm H}]_s$) and in the solar 
abundances (which affects the conversion between $[{\rm Fe}/{\rm H}]_s$ and $Z_s$). Note that once we have large enough sample, we can alleviate 
the problem of poor precision of $Z_i$, estimate the impact of the systematic uncertainty in $Z_i$, and also take more conservative approach 
and restrict the stars according to $\delta Y < 0.040$ to gain more confidence in the estimates of $Y_p$ and $\Delta Y / \Delta Z$. 

The ``simple" models of galactic chemical evolution predict a linear relationship between $Y_i$ and $Z_i$ \citep[see e.g.][]{hacy76}, however 
the reality may be more complex. Recent studies indicate towards a mildly quadratic relationship between $Y_i$ and $Z_i$ \citep{west13}. It is 
currently not possible to constrain a quadratic relationship given the small sample size and large errorbars, however once the sample size grows 
bigger in the future, such studies would be very interesting.

\section{Summary}
\label{summ}
We used the helium glitch signature to estimate the surface helium abundance of the stars in the LEGACY sample. We found that $Y_s$ for the 
low-mass stars can be determined reliably only if the number of the observed oscillation frequencies and their precision are  good enough to 
determine the average amplitude of the helium signature reliably. On the other hand, for the high-mass stars the larger uncertainties on the 
oscillation frequencies and hence the frequency ratios make $Y_s$ determination difficult. We found 38 stars in the LEGACY sample for which $Y_s$ 
can be determined reliably. 

We extracted the glitch signatures from both, the oscillation frequencies and its second differences (Methods A and B), to quantify the systematic 
uncertainties on $Y_s$ associated with the treatment of the background smooth component of the oscillation frequency. The two methods provide 
$Y_s$ within $1\sigma$ for all the 38 stars. 

We calibrated the observed average amplitude against the corresponding amplitude obtained from the model frequencies with different $Y_s$ to 
estimate the surface helium abundance. Since the calibration involved the stellar models which are known to be uncertain, we used four different 
sets of models -- one each using the MESA and GARSTEC and two using the YREC -- with different input physics to quantify the uncertainties on 
$Y_s$ associated with the uncertainties in the stellar physics. The different estimates of $Y_s$ obtained in this manner agree within $1\sigma$ 
for most of the stars. However, a systematic difference of about 0.02 was noted in $Y_s$ when obtained using the diffusion and non-diffusion 
models.

We used the measurements of the surface abundances together with the settling predicted by the stellar models to compute the initial abundances. 
The initial abundances were used to derive the primordial helium abundance, $Y_p = 0.244 \pm 0.019$, and the enrichment ratio, 
$\Delta Y / \Delta Z = 1.226 \pm 0.849$. Currently in the {\it Kepler} era, the uncertainties are large because of the small sample size, and 
we hope to be able to determine these quantities with much higher precision in the PLATO era.

\section*{Acknowledgements}
Funding for the Stellar Astrophysics Centre is provided by The Danish National Research Foundation (Grant agreement no.: DNRF106). We thank the 
anonymous referee for the constructive feedbacks. We are grateful to H.~M.~Antia for the discussions we had in the beginning of the project and for 
his comments on an earlier version of the manuscript. KR, AM and PR acknowledge support from the NIUS program of HBCSE (TIFR). SB is partially 
supported by NSF grant AST-1514676 and NASA grant NNX16AI09G. VSA acknowledges support from VILLUM FONDEN (Research Grant 10118). MNL acknowledges 
the support of The Danish Council for Independent Research | Natural Science (Grant DFF-4181-00415).


\appendix

\section{Different measures of the amplitude of the helium glitch signature}
\label{adv}
\begin{figure}
\includegraphics[width=\columnwidth]{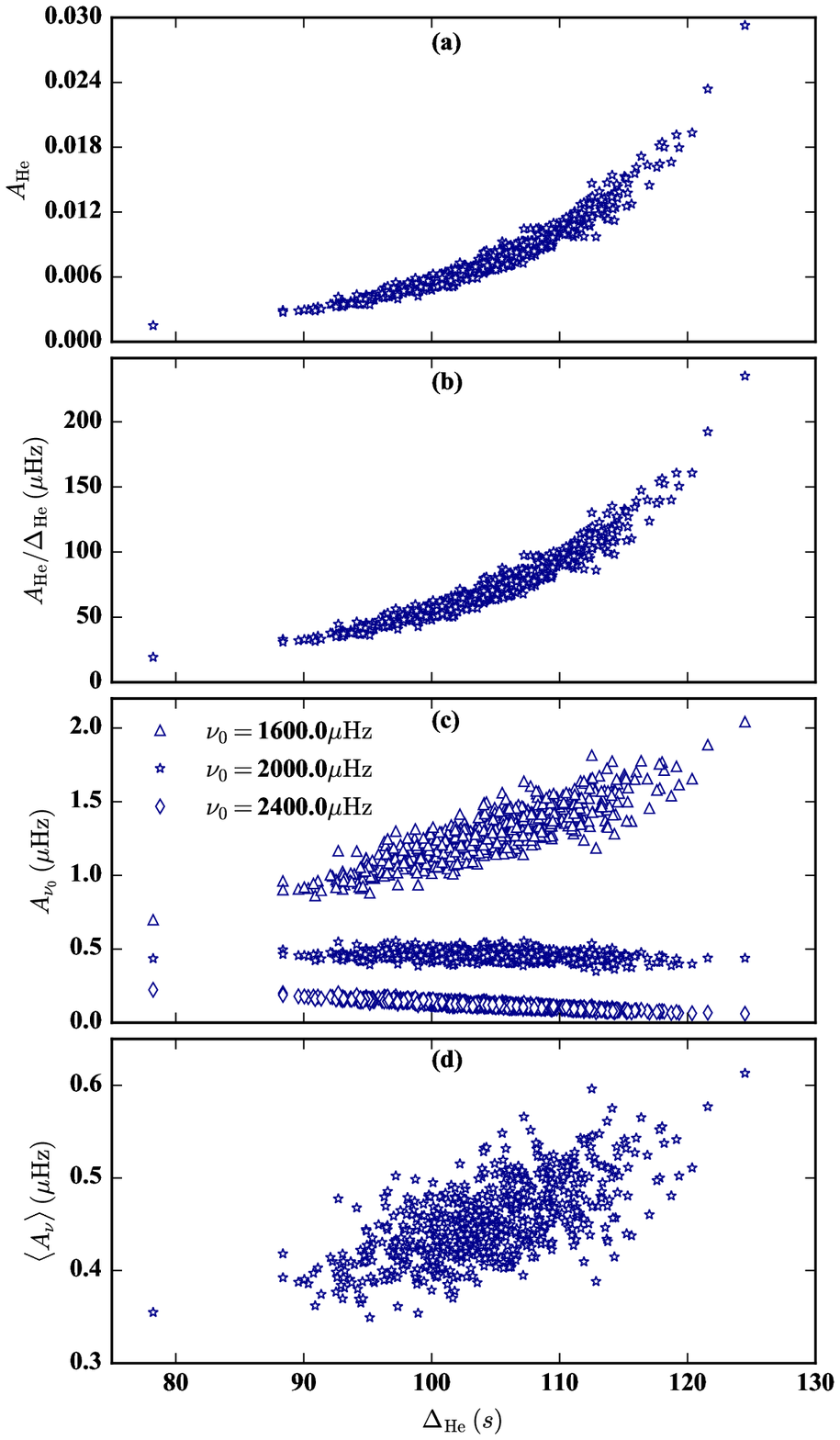}
\caption{Different measures of the amplitude as a function of $\Delta_{\rm He}$ for the different realizations of 16 Cyg A data. The three different 
types of points in panel (c) represent the amplitudes at the three different reference frequencies.}
\label{fig8}
\end{figure}

In principle, we can use any of the four measures of the amplitude discussed in Section~\ref{amp} for the calibration. In practice, however, 
$A_{\rm He}$ and $A_{\rm He}/\Delta_{\rm He}$ are poorly determined due to the uncertainties in the observed frequencies. Both of 
these quantities are tightly correlated with $\Delta_{\rm He}$, and the resulting trade-off leads to the large observational errorbar on them. For 
an example in Figure~\ref{fig8}, we show the scatter in $A_{\rm He}$, $A_{\rm He}/\Delta_{\rm He}$, $A_{\nu_0}$ and $\langle A_{\nu} \rangle$ as a 
function of $\Delta_{\rm He}$ obtained by fitting the different realizations of 16 Cyg A data using the Method A. The spread along an axis represents
the observational uncertainty on the corresponding quantity. Note a factor of 4 variation of $A_{\rm He}$ and $A_{\rm He}/\Delta_{\rm He}$ in panels 
(a-b). The $A_{\nu_0}$ is also correlated with $\Delta_{\rm He}$ for an arbitrary choice of $\nu_0$, but its variation is only a factor of 2 in the 
worse case as seen in panel (c). The correlation changes from being positive to negative as $\nu_0$ increases, and we can choose a value of $\nu_0$ 
which corresponds to approximately zero correlation and get the corresponding most precise $A_{\nu_0}$. Instead of determining the most appropriate 
value of $\nu_0$ and using the corresponding $A_{\nu_0}$ for the calibration, we used average amplitude, $\langle A_{\nu} \rangle$. As can be seen 
in panel (d), this choice is better than the choices of $A_{\rm He}$ and $A_{\rm He}/\Delta_{\rm He}$ and also $A_{1600}$ and $A_{2400}$, but 
slightly worse than $A_{2000}$.

\section{Impact of the surface term}
\label{surface}

\begin{figure}
\includegraphics[width=0.95\columnwidth]{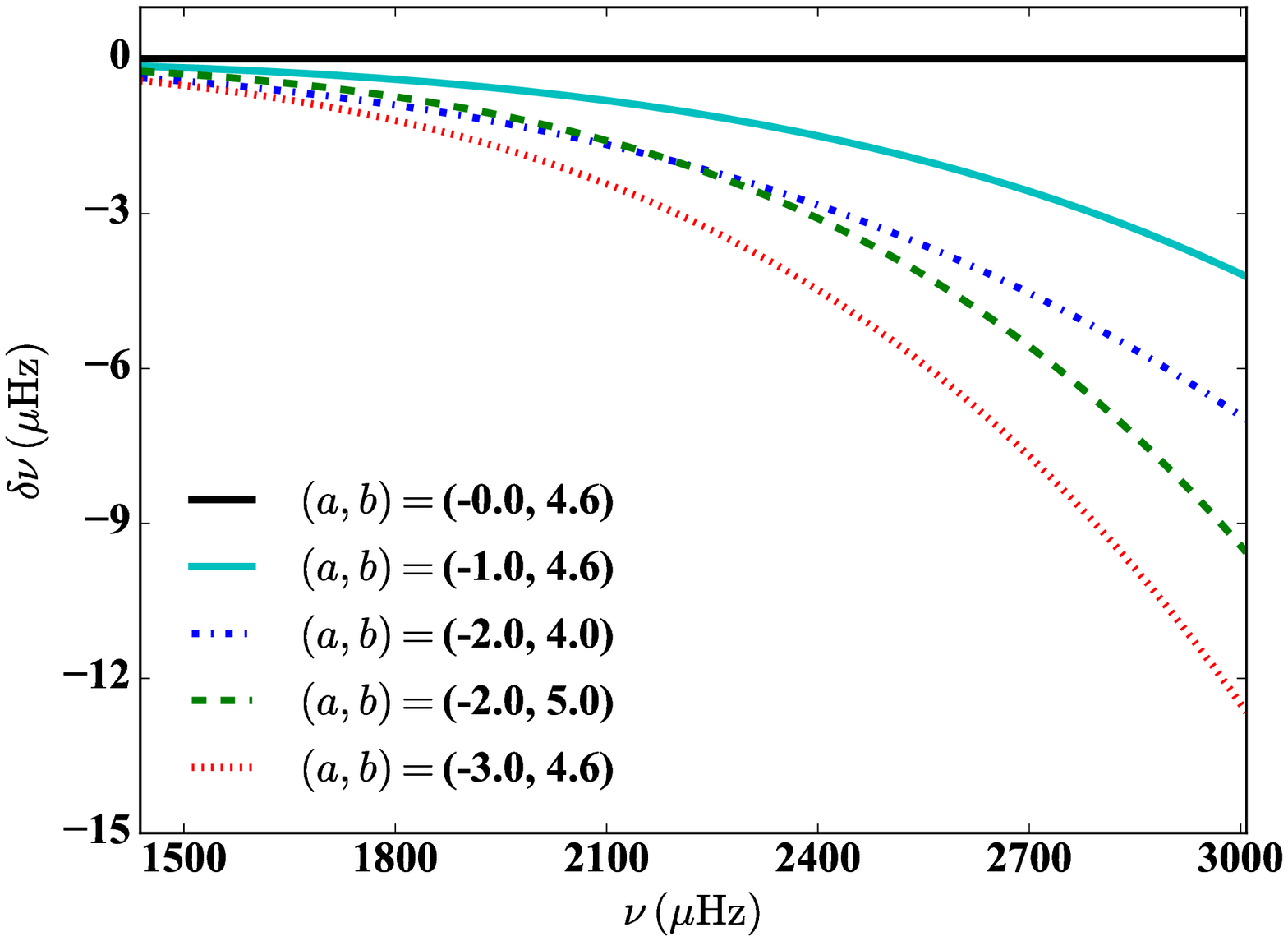}
\caption{Corrections applied to the best-fitting model frequencies of 16 Cyg A. The different lines correspond to the different values of $a$ and 
$b$ (see the text). The horizontal line corresponds to the original model frequencies.} 
\label{fig9}
\end{figure}

\begin{table*}
\centering
\scriptsize
\caption{Fitted parameters for the corrected model frequencies of 16 Cyg A. The $\sigma_{\rm stat}$ is the mean of the negative and positive 
uncertainties obtained using the Monte-Carlo simulation for the corresponding observed parameter.}
\label{tab5}
\begin{tabular}{crrrrrrrrrrr}
\hline\hline
(a, b)  &  \multicolumn{5}{c}{Method A}  &  &  \multicolumn{5}{c}{Method B}\\
\cline{2-6} \cline{8-12}\\
  &  $\langle A_{\rm He} \rangle$  &  $\Delta_{\rm He}$  &  $\tau_{\rm He}$  &  $\langle A_{\rm CZ} \rangle$  &  $\tau_{\rm CZ}$  &
  &  $\langle A_{\rm He} \rangle$  &  $\Delta_{\rm He}$  &  $\tau_{\rm He}$  &  $\langle A_{\rm CZ} \rangle$  &  $\tau_{\rm CZ}$\\
  &  ($\mu$Hz)  &  (s)  &  (s)  &  ($\mu$Hz)  &  (s)  &  &  ($\mu$Hz)  &  (s)  &  (s)  &  ($\mu$Hz)  &  (s)\\
\hline
(-0.0, 4.6)  &  0.4043  &  107.34  &  890.48  &  0.0732  &  3022.2  &  &  0.4190  &  105.55  &  911.28  &  0.0725  &  3015.9\\
(-1.0, 4.6)  &  0.4035  &  107.73  &  889.20  &  0.0729  &  3026.4  &  &  0.4209  &  106.05  &  908.32  &  0.0721  &  3019.4\\
(-2.0, 4.0)  &  0.4039  &  107.76  &  889.52  &  0.0729  &  3030.0  &  &  0.4220  &  106.12  &  908.37  &  0.0721  &  3022.9\\
(-2.0, 5.0)  &  0.4088  &  108.51  &  886.38  &  0.0730  &  3030.6  &  &  0.4316  &  106.98  &  902.26  &  0.0721  &  3022.1\\
(-3.0, 4.6)  &  0.4098  &  108.54  &  886.22  &  0.0729  &  3034.6  &  &  0.4337  &  107.07  &  901.56  &  0.0720  &  3025.9\\
$\sigma_{\rm stat}$  &  0.0430  &  5.89  &  33.30  &  0.0158  &  71.7  &  &  0.0395  &  6.05  &  27.30  &  0.0147  &  71.1\\
\hline
\end{tabular}
\end{table*}

The observed and best-fitting model frequencies are systematically different due to the differences in the near-surface layers of the star and 
the model. This difference is known as the ``surface term". The helium ionization zones in the solar-type stars lie deeper where the convection 
is adiabatically stratified, and hence we can expect 1D stellar models to reproduce these layers. Therefore, the helium glitch signature in
the best-fitting model frequencies should reproduce the corresponding observed signature. However, since the observed and best-fitting model 
frequencies are different, it is not clear while fitting that the additional surface term in the model frequencies goes completely to the smooth 
component, $\nu_{\rm smooth}$ in case of Method A and $\delta^2\nu_{\rm smooth}$ in case of Method B, without affecting the glitch signatures.

To estimate the effect of the surface term on the glitch signatures, we corrected the best-fitting model frequencies of 16 Cyg A following the 
power-law correction of \citet{kjel08},
\begin{equation}
\delta\nu = a \left(\frac{\nu}{\nu_{\rm ref}}\right)^b,
\end{equation}
where $a$ and $b$ are constants, and $\nu_{\rm ref} = 2200$ $\mu$Hz is a reference frequency. Figure~\ref{fig9} shows the five different corrections 
corresponding to the five arbitrary choices of $a$ and $b$. We fitted the resulting frequencies using the Methods A and B, and the fitted parameters 
are listed in Table~\ref{tab5}. Clearly the impact of the surface term is smaller than the impact of the statistical uncertainty on the observed 
frequencies from the {\it Kepler} satellite. The reason is that the observational error is a rapidly varying function of the frequency, like the 
glitch signatures themselves, and interfere with them. On the other hand, since the surface term is a slowly varying function of frequency, it 
contributes to the smooth component, $\nu_{\rm smooth}$ in case of Method A and $\delta^2\nu_{\rm smooth}$ in case of Method B, without affecting 
the glitch signatures in any significant way.

\bsp
\label{lastpage}
\end{document}